\documentclass[12pt]{iopart}
\usepackage{iopams}
\usepackage{graphicx}

\begin{document}
 
\title[Cosmic Topology of Polyhedral Double-Action Manifolds]
{Cosmic Topology of Polyhedral Double-Action Manifolds}
\author{R.\ Aurich and S.\ Lustig}
 
\address{Institut f\"ur Theoretische Physik, Universit\"at Ulm,\\
Albert-Einstein-Allee 11, D-89069 Ulm, Germany}

\begin{abstract}
A special class of non-trivial topologies of the spherical space ${\cal S}^3$
is investigated with respect to their cosmic microwave background (CMB)
anisotropies.
The observed correlations of the anisotropies on the CMB sky possess
on large separation angles surprising low amplitudes
which might be naturally be explained by models of the Universe
having a multiconnected spatial space.
We analysed in CQG 29(2012)215005 the CMB properties of prism
double-action manifolds that are generated by a binary dihedral group
$D^\star_p$ and a cyclic group $Z_n$ up to a group order of 180.
Here we extend the CMB analysis to polyhedral double-action manifolds
which are generated by the three binary polyhedral groups
($T^\star$, $O^\star$,  $I^\star$) and a cyclic group $Z_n$
up to a group order of 1000.
There are 20 such polyhedral double-action manifolds.
Some of them turn out to have even lower
CMB correlations on large angles than the Poincar\'e dodecahedron.
\end{abstract}


\submitto{\CQG}

\section{Introduction.}
\label{sec:intro}

The $\Lambda$CDM concordance cosmological model describes nearly all 
cosmological observations very successfully.
Among the few exceptions is the observation of 
the COBE team \cite{Hinshaw_et_al_1996}
that the fluctuations in the cosmic microwave background (CMB)
are nearly uncorrelated on large angular scales $\vartheta \gtrsim 60^\circ$.
This surprising result is confirmed by the WMAP team \cite{Spergel_et_al_2003}
and further discussed in \cite{Aurich_Janzer_Lustig_Steiner_2007,%
Copi_Huterer_Schwarz_Starkman_2008,Copi_Huterer_Schwarz_Starkman_2010}
with respect to the $\Lambda$CDM concordance model.
In \cite{Efstathiou_Ma_Hanson_2009} it is argued that there
is no significant deviant behaviour from the $\Lambda$CDM model
if the uncertain parts in the CMB map are suitably reconstructed
from the less uncertain regions.
However, the reconstruction algorithm is analysed by
\cite{Aurich_Lustig_2010,Copi_Huterer_Schwarz_Starkman_2011}
showing that this method does not lead to a robust measure
of the true CMB sky and the use of masked sky maps is to be preferred.
It is concluded in \cite{Copi_Huterer_Schwarz_Starkman_2011}
that the ``lack of large-angle correlation,
particularly on the region of the sky outside the Galaxy,
remains a matter of serious concern.''

In this paper we try to explain the uncorrelated CMB fluctuations
on large scales by relaxing the assumption of the concordance model
that the Universe possesses a simply connected spatial topology.
Instead, non-trivial topologies are assumed for the spatial 3-manifold,
i.\,e.\ multiconnected spaces,
which can lead to a suppression of CMB correlations on angles
corresponding the topological length scale.
The simply connected space of the $\Lambda$CDM concordance model
possesses one of the three curvature properties:
Euclidean for the ${\cal E}^3\equiv\mathbb{R}^3$,
spherical for the ${\cal S}^3$, or
hyperbolic for the ${\cal H}^3$ depending on the total density
$\Omega_{\hbox{\scriptsize tot}}$.
These three simply connected spaces are considered as the
universal cover which is tessellated by a deck group $\Gamma$
into cells which are identified.
The size of such a cell defines the topological length scale.
For an introduction into the topic of cosmic topology, see
\cite{Lachieze-Rey_Luminet_1995,Luminet_Roukema_1999,Levin_2002,%
Reboucas_Gomero_2004,Luminet_2008,Fujii_Yoshii_2011}.
Below the topological length scale the properties of the concordance model
are not altered since the cosmological parameters of the
$\Lambda$CDM concordance model are used,
and the local physics is unchanged.
For example, possible non-Gaussian features in the CMB are the same
as predicted by the $\Lambda$CDM concordance model
\cite{Monteserin_Barreiro_Sanz_Martinez-Gonzalez_2005}.
It is shown in \cite{Aurich_Janzer_Lustig_Steiner_2010}
that the fine structure of the CMB fluctuations for the 
$\Lambda$CDM concordance model and for the 3-torus topology
cannot be distinguished experimentally due to the same local physics.

We investigate the statistical properties
of the CMB anisotropies on large separation angles
that arise in polyhedral double-action manifolds.
These models are not studied in the literature and thus,
their CMB properties are unknown.
As discussed below, the considered polyhedral double-action manifolds
derive from parent manifolds having one of the most severe suppressions
of CMB correlations on large scales.
This motivates the investigation of polyhedral double-action manifolds
since one can hope that they inherit the suppression.
These models require a spherical 3-space ${\cal S}^3$
but we mostly restrict our analysis to almost flat spaces corresponding to
a total density $\Omega_{\hbox{\scriptsize tot}}$ in the range
$\Omega_{\hbox{\scriptsize tot}}=1.001,\dots,1.05$.
The multiconnected spaces that exist in the spherical 3-space ${\cal S}^3$
can be classified with respect to three categories of
spherical 3-manifolds
as described in \cite{Gausmann_Lehoucq_Luminet_Uzan_Weeks_2001}.
The criterion is based on the kind of two subgroups $R$ and $L$
which generate the deck group $\Gamma$
which in turn defines the spherical 3-manifold.
The subgroups $R$ and $L$ act as pure right-handed and
left-handed Clifford translations, respectively.
The first category consists of the single-action manifolds
in which only one of the subgroups $R$ and $L$ acts non-trivially.
The double-action manifolds, the second category, require that
both subgroups $R$ and $L$ are non-trivial,
such that each element of the subgroup $R$
is combined with each element of the subgroup $L$.
The third category, the linked-action manifolds, are similar to the
second one, except that there are rules specifying which elements of $R$
and $L$ can be combined such that a manifold is obtained
instead of an orbifold.
For more details on the three categories,
see \cite{Gausmann_Lehoucq_Luminet_Uzan_Weeks_2001}.

The single-action manifolds are the simplest
with respect to an analysis of the statistical CMB properties,
since they are independent of the position of the
CMB observer within the manifold.
Such manifolds are called homogeneous.
This contrasts to the other two categories
where the ensemble average of the CMB statistics depends on the
observer position, in general, and a much more involved analysis is required
for these inhomogeneous manifolds.

The aim of this paper is to close a gap
that is left by our previous publications
\cite{Aurich_Lustig_2012b,Aurich_Lustig_2012c}
which cover some of the possible double-action manifolds.
A survey of lens spaces $L(p,q)$ is presented in \cite{Aurich_Lustig_2012b}.
The lens spaces $L(p,q)$ have the amazing property
that they have members in all three categories.
While the spaces $L(p,1)$ are single-action manifolds,
the lens spaces $L(m n,q)$ which are generated by $R=Z_m$ and $L=Z_n$
with $m$ and $n$ relatively prime, are double-action manifolds.
The remaining lens spaces belong to the linked-action manifolds
so that members of all three categories are studied
in \cite{Aurich_Lustig_2012b}.
This study leads to the result that lens spaces $L(p,q)$
with $q\simeq 0.28 p$ or  $q\simeq 0.38 p$ possess a pronounced
suppression of CMB correlations on large angular scales
compared to other lens spaces.
The prism double-action manifolds, which are generated by
a binary dihedral group $R=D^\star_p$ and a cyclic group $L=Z_n$,
are investigated in \cite{Aurich_Lustig_2012c}, and at least
three promising spaces are found.
In the notation of \cite{Aurich_Lustig_2012c},
the prism double-action manifolds are called $DZ(p,n)$
where the letters indicate the subgroups $R$ and $L$, and
$p$ and $n$ are the group orders of $D^\star_p$ and $Z_n$.
Three prism double-action manifolds with a remarkable
large-scale CMB suppression are $DZ(8,3)$, $DZ(16,3)$, and $DZ(20,3)$.
Because of these encouraging results, the question emerges
whether there are further interesting double-action manifolds.
The double-action manifolds not covered in \cite{Aurich_Lustig_2012b}
and \cite{Aurich_Lustig_2012c} are those generated by
one of the three binary polyhedral groups $R = T^\star$, $O^\star$ or $I^\star$
and a cyclic group $L=Z_n$.
For these spaces we introduce the notation $TZ(24,n)$, $OZ(48,n)$,
and $IZ(120,n)$.
Thus, this paper is devoted to these spaces
in order to close the gap with respect to the
CMB properties of polyhedral double-action manifolds.
We investigate all 20 polyhedral double-action manifolds
which exist up to the group order 1000.

The polyhedral double-action manifolds can be considered as a
dissection of one of the three polyhedral spaces with respect
to a cyclic group.
The three polyhedral spaces belong to the single-action spaces
and are thus homogeneous.
They are well studied in several previous papers
starting with \cite{Luminet_Weeks_Riazuelo_Lehoucq_Uzan_2003}
which analyses the Poincar\'e dodecahedral topology
that is the binary icosahedral space ${\cal I}$.
A strong suppression of CMB correlations on large angular scales is found
for this space at $\Omega_{\hbox{\scriptsize tot}}\simeq 1.02$.
This result is confirmed in \cite{Aurich_Lustig_Steiner_2004c}
by using a much larger set of eigenfunctions for the computation
of the CMB statistics.
Further studies concerning this model can be found in
\cite{Roukema_et_al_2004,Gundermann_2005,Aurich_Lustig_Steiner_2005a,%
Aurich_Lustig_Steiner_2005b,Lustig_2007,%
Key_Cornish_Spergel_Starkman_2007,Niarchou_Jaffe_2007,%
Lew_Roukema_2008,Roukema_et_al_2008a,Roukema_Kazimierczak_2011}.
In \cite{Gundermann_2005,Aurich_Lustig_Steiner_2005a,Niarchou_Jaffe_2007}
the statistical CMB analysis is extended to the
binary tetrahedral space ${\cal T}$ and the binary octahedral space ${\cal O}$.
The central result of \cite{Aurich_Lustig_Steiner_2005a} is that all three
polyhedral spaces lead to a significant suppression of large-scale correlations
described by the $S$ statistics of a factor of
$\sim 0.11$ compared to the simply connected spherical 3-space ${\cal S}^3$.
This factor is achieved at $\Omega_{\hbox{\scriptsize tot}}\simeq 1.07$,
$\Omega_{\hbox{\scriptsize tot}}\simeq 1.04$, and
$\Omega_{\hbox{\scriptsize tot}}\simeq 1.02$ for the spaces
${\cal T}$,  ${\cal O}$, and ${\cal I}$, respectively.
In the following we analyse the statistical properties on large separation
angles $\vartheta$ of the polyhedral double-action manifolds in order
to address the question
how strong these spaces suppress the CMB correlations in terms
of the $S$ and $I$ statistics defined below
in eqs.\,(\ref{Eq:S_statistic_60}) and (\ref{Eq:I_measure}).
Since they are based on the three polyhedral spaces with their very
low values of the $S$ statistics, they also could yield promising models
for the description of our Universe.

The polyhedral double-action manifolds are generated by
a cyclic subgroup $L=Z_n$ and one of the three
binary polyhedral groups $R = T^\star$, $O^\star$, and $I^\star$,
where the cyclic groups $Z_n$ have to fulfil 
$\hbox{gcd}(24,n)=1$, $\hbox{gcd}(48,n)=1$, and $\hbox{gcd}(120,n)=1$,
respectively. 
The generator $g_l=({\bf 1},g_b)$ of the cyclic group $Z_n$ is given by
\begin{equation}
\label{Def:Z_n}
g_b \; = \;
\hbox{diag}(e^{+2\pi\hbox{\scriptsize i}/n},e^{-2\pi\hbox{\scriptsize i}/n})
\hspace{10pt}.
\end{equation}
The binary polyhedral groups $R = T^\star$, $O^\star$, and $I^\star$ have 
two generators $g_{r1}=(g_{a1 },{\bf 1})$ and $g_{r2}=(g_{a2 },{\bf 1})$. 
These two generators can be described by  
\begin{equation}
\label{Def:poly_group}
g_{ak} \; = \;
\left(\begin{array}{cc}
\cos(\tau_k)-\hbox{i}\sin(\tau_k)\cos(\theta_k) &
-\hbox{i}\sin(\tau_k)\sin(\theta_k)e^{-\hbox{\scriptsize i}\phi_k}\\ 
-\hbox{i}\sin(\tau_k)\sin(\theta_k)e^{\hbox{\scriptsize i}\phi_k}&
\cos(\tau_k)+\hbox{i}\sin(\tau_k)\cos(\theta_k) 
\end{array}\right)
\hspace{6pt} 
\end{equation}
using the spherical coordinates $(\tau_k, \theta_k, \phi_k)$, $k=1,2$.
The values of $\tau_k$, $\theta_k$ and $\phi_k$ given in 
table \ref{Tab:generators_poly_spherical} determine the representation of
the groups  $T^\star$, $O^\star$, and $I^\star$.   
\begin{table}[!htbp]
\centering
 \begin{tabular}{|c|c|c|}
 \hline
  group $R$ & ($\tau_1$, $\theta_1$, $\phi_1$) & ($\tau_2$, $\theta_2$, $\phi_2$)\\
 \hline
 $T^\star$ & $(\frac{\pi}{3}, 0, 0)$&$(\frac{\pi}{3}, \arccos\big(\frac{1}{3}\big), 0)$\\
 \hline
 $O^\star$ & $(\frac{\pi}{4}, 0, 0)$&$(\frac{\pi}{3}, \arccos\big(\frac{1}{\sqrt 3}\big), 0)$\\
 \hline
 $I^\star$ & $(\frac{\pi}{5}, 0, 0)$&$(\frac{\pi}{5}, \arccos\big(\frac{1}{\sqrt 5}\big), 0)$\\
 \hline
 \end{tabular}
 \caption{\label{Tab:generators_poly_spherical}
 These values of $(\tau_1, \theta_1, \phi_1)$ and  $(\tau_2, \theta_2, \phi_2)$
 determine the two generators in eq.\,(\ref{Def:poly_group}) for the 
 binary polyhedral groups  $T^\star$, $O^\star$, and $I^\star$. 
}
\end{table}

Although the central topic of this paper concerns the
correlation of the CMB fluctuations on large angular scales,
some remarks on the circles-in-the-sky (CITS) signature are in order
which serves as a topological test \cite{Cornish_Spergel_Starkman_1998b}.
The CITS test requires a full CMB sky survey and has been applied
to different sky maps derived from the WMAP mission.
The first year CMB data are analysed with respect to
nearly back-to-back circle pairs by 
\cite{Cornish_Spergel_Starkman_Komatsu_2003,Key_Cornish_Spergel_Starkman_2007}
and no significant signature was found,
whereas a search for the Poincar\' e dodecahedral space,
being a single action manifold, yields a tentative signal
\cite{Roukema_et_al_2004}.
It is shown in \cite{Aurich_Janzer_Lustig_Steiner_2007}
that the error in the CMB signal has to be significantly lower than
50$\mu\hbox{K}$ in order to get a CITS signal.
It is hard to obtain a statement about the size of the error
in the heavily processed WMAP data leading to the maps
used for the CITS searches.
The constraint to nearly back-to-back circle pairs is investigated in
\cite{Mota_Reboucas_Tavakol_2010,Mota_Reboucas_Tavakol_2011}
where the probability for the deviation from the back-to-back orientation
is studied.
The seven year WMAP data are analysed by \cite{Bielewicz_Banday_2011}
again for the special case of back-to-back circles,
and no topological signature is detected.
A complete CITS search without the back-to-back restriction
is carried out in \cite{Vaudrevange_Starkman_Cornish_Spergel_2012}
using the WMAP seven year data.
Several signatures are found, but they are all ascribed to foreground sources,
so that the paper concludes that no hint for a non-trivial topology is found.
Since no statement on the accuracy of the CMB signal is made,
one cannot exclude the possibility that a possible CITS signal is
swamped by foreground sources which can even produce spurious signals.
In order to reduce the computer time, the analysis of
\cite{Vaudrevange_Starkman_Cornish_Spergel_2012}
uses a search grid for the screening of circle pairs
that is coarser than that of the CMB map.
Our preliminary investigations show that the probability for missing
circle pairs increases by such an algorithm.
For this reason topologies with few circle pairs have a high probability
to get missed in this way.
Since these results are devoted to a future publication,
we turn to the CMB correlations now.

\section{Eigenmodes on Polyhedral Double-Action Manifolds}
\label{sec:eigenmode}

The CMB analysis on spherical manifolds requires
the computation of the eigenmodes of the Laplace-Beltrami operator $\Delta$
expanded with respect to the spherical basis $| j; l, m \rangle$.
The starting point is the abstract basis $| j; m_a, m_b \rangle$ with
$2j\in\mathbb{N}_0$, $|m_a| \leq j$, and $|m_b| \leq j$,
which can be written as a product
\begin{equation}
\label{Eq:SO4_Basis}
| j; m_a, m_b \rangle \; := \;
|j,m_a\rangle \, |j,m_b\rangle \in \hbox{SO}(4,\mathbb{R})
\hspace{10pt} ,
\end{equation}
in an eigenbasis for the abstract generators
$\vec J_a = (J_{ax},J_{ay},J_{az})$ and $\vec J_b = (J_{bx},J_{by},J_{bz})$
of two Lie algebras on $\hbox{SU}(2,\mathbb{C})$.
The number $j$ is related to the eigenvalue $E_j$ of $-\Delta$ by
$E_j = 4j(j+1) = \beta^2-1$, where $\beta = 2j+1$ is the wave number.

The eigenmodes of $\Delta$ have to satisfy the periodic boundary conditions
imposed by the deck group.
The eigenstates of the polyhedral double-action manifolds can be obtained
by considering only the generators of the subgroups $R$ and $L$.
The generator (\ref{Def:Z_n}) of the subgroup $L=Z_n$ acts as
$U_{g_{l}} = e^{\hbox{\scriptsize i} \frac{4\pi}{n}J_{bz}}$ on $| j; m_a, m_b \rangle$
which leads to the selection rule
\begin{equation}
\label{Eq:selection_rule_mb}
2 m_b \; \equiv \; 0 \hbox{ mod } n
\hspace{10pt} .
\end{equation}
A further restriction is obtained by the action
$U_{g_{r1}}=e^{\hbox{\scriptsize i} \frac{2\pi}{N}J_{az}}$
of the first generator $g_{r1}$, eq.\,(\ref{Def:poly_group}),
of the binary polyhedral group on $| j; m_a, m_b \rangle$
which requires for $m_a$ the selection rule
\begin{equation}
\label{Eq:selection_rule_ma}
m_a\equiv 0 \hbox{ mod } N
\end{equation}
with $N=3$ for $T^\star$, $N=4$ for $O^\star$, and $N=5$ for $I^\star$.
The action of the second generator of the binary polyhedral group cannot
be incorporated by such a simple selection rule.
This contrasts to the corresponding relations for $L(p,q)$ and $DZ(p,n)$
which can be analytically solved leading to the results stated
in \cite{Aurich_Lustig_2012a} and \cite{Aurich_Lustig_2012c}.
Thus, the eigenstates have to be expressed by the ansatz
\begin{equation}
\label{Eq:Ansatz_red_Basis}
| j; s, m_b \rangle \; = \;
\sum_{m_a\equiv 0 \;\hbox{\scriptsize mod}\;N} a^s_{m_a}| j; m_a, m_b \rangle
\hspace{10pt} \hbox{ with } \hspace{10pt}
2 m_b \equiv 0 \hbox{ mod } n
\hspace{10pt} .
\end{equation}
The coefficients $a^s_{m_a}$ have to be determined from
the system of equations obtained from the boundary conditions of the
second generator $g_{r2}$
where the solutions $a^s_{m_a}$ are independent of $m_b$.
The index $s$ counts the linearly distinct solutions
(\ref{Eq:Ansatz_red_Basis}) of that system of equations.

With respect to the spherical coordinates $(\tau,\theta,\phi)$
the eigenmodes are  given by 
$\psi^{\, \cal M}_{j;s,m_b}(\tau,\theta,\phi):=\langle \tau,\theta,\phi| j; s, m_b \rangle$.
Considering the action of the generator $g_{r2}$ on
the eigenmode $\psi^{\, \cal M}_{j;s,m_b}(\tau,\theta_2,\phi_2)$ with the values of
$\theta_2$ and $\phi_2$ given in table \ref{Tab:generators_poly_spherical},
one obtains the transformed eigenmode
$\psi^{\, \cal M}_{j;s,m_b}(\tau+\tau_2,\theta_2,\phi_2)$
in terms of the coefficients $a^s_{m_a}$.
This leads with
$\psi^{\, \cal M}_{j;s,m_b}(\tau,\theta_2,\phi_2)-
\psi^{\, \cal M}_{j;s,m_b}(\tau+\tau_2,\theta_2,\phi_2)=0$
to a system of equations
whose solution yields the coefficients $a^s_{m_a}$.
This system of equations has to be solved numerically
as outlined in \ref{sec:eigenmodes_polyhedral}
and \ref{sec:eigenmodes_polyhedral_da},
see also \cite{Gundermann_2005,Lustig_2007}.
To each eigenvalue $E_j$ there exists $r^{\cal M}(\beta)$ eigenmodes 
which we denote as $|j,i \rangle$, where $i$ counts the degenerated modes.
The wave number spectrum $\beta$ as well as the corresponding
multiplicities $r^{\cal M}(\beta)$ are given in table 1 in
\cite{Aurich_Lustig_2012c}.

For the CMB analysis the expansion of the eigenmodes in the spherical basis
$| j; l, m \rangle$ is required with respect to the observer position.
To specify this position, the transformation $t$ is introduced as
\begin{equation}
\label{Eq:coordinate_t_rho_alpha_epsilon}
t(\rho,\alpha,\epsilon) \; = \;
\left( \begin{array}{cc}
\cos(\rho)\, e^{+\hbox{\scriptsize i}\alpha} &
\sin(\rho)\, e^{+\hbox{\scriptsize i}\epsilon} \\
-\sin(\rho)\, e^{-\hbox{\scriptsize i}\epsilon} &
\cos(\rho)\, e^{-\hbox{\scriptsize i}\alpha}
\end{array} \right)
\end{equation}
with $\rho \in [0,\frac{\pi}{2}]$, $\alpha, \epsilon \in [0,2\pi]$.
The transformation $t$ is defined as right multiplication.
Applying this transformation to the position of the observer at
the origin of the given coordinate system
generates a set of new observer positions
parameterised by $\rho$, $\alpha$, and $\epsilon$.
The expansion of the eigenmodes with respect to the new observer position
is found to be (see \ref{sec:eigenmodes_polyhedral_da})
\begin{eqnarray}
D(t^{-1})|j,i \rangle
\nonumber =
\sum_{l=0}^{2j}\sum_{m=-l}^{l} \xi^{j,i}_{lm}({\cal M};t)\,
|j; l, m \rangle
\;\;\\
\label{Eq:eigenfunction_TZ_IZ_OZ_exp_sph}
\xi^{j,i}_{lm}({\cal M};t)
=\sum_{\tilde{m}_b}\langle jm_aj\tilde{m}_b|lm\rangle a^{s(i)}_{m_a} \,
D^{\,j}_{\tilde{m}_b,m_b(i)}(t^{-1})\\
\nonumber
\hbox{with} \hspace{5pt} m_a + \tilde{m}_b = m  \hspace{5pt} , \hspace{5pt}
m_a\equiv 0 \;\hbox{mod}\;N \hspace{5pt}\hbox{and} 
\hspace{5pt}2\,m_b(i)\equiv 0 \;\hbox{mod}\;n
\hspace{10pt} .
\end{eqnarray}
The values of $N$ are $N=3$ for $TZ(24,n)$, $N=4$ for $OZ(48,n)$,
and $N=5$ for $IZ(120,n)$.
Furthermore, $\langle j m_a j \tilde{m}_b | lm \rangle$ are the
Clebsch-Gordan coefficients \cite{Edmonds_1964}, and
$D^{\,j}_{\tilde{m}_b,m_b}(t)$ are the Wigner polynomials
\begin{equation}
\label{Eq:D_function_rho_alpha_epsilon}
\nonumber
D^{\,j}_{\tilde{m}_b,m_b}(t)
\; := \; \langle j, \tilde{m}_b |D(t)| j,  m_b \rangle
\; = \; 
e^{\hbox{\scriptsize i}\,(\alpha + \epsilon)\,\tilde{m}_b}
d^{\,j}_{\tilde{m}_b, m_b}(2 \rho)
e^{\hbox{\scriptsize i}\,(\alpha - \epsilon)\, m_b}
\hspace{10pt} .
\end{equation}
With the coefficients $\xi^{j,i}_{lm}({\cal M};t)$
the CMB statistics can be computed since they allow the
calculation of the multipole moments
\begin{equation}
\label{Eq:Cl_ensemble}
C_l :=
\frac{1}{2l+1}\sum_{m=-l}^l\left\langle\left|a_{lm}\right|^2\right\rangle
=
\sum_{\beta}\frac{ T_l^2(\beta) \; P(\beta)}{2l+1}\sum_{m=-l}^{l}
\sum_{i=1}^{r^{\cal M}(\beta)} \left|\xi^{\beta,i}_{lm}({\cal M};t)\right|^2
\end{equation}
as shown in \cite{Aurich_Lustig_2012a,Aurich_Lustig_2012c}.
The initial power spectrum is
$P(\beta)\sim 1/(E_{\beta}\,\beta^{2-n_{\hbox{\scriptsize s}}})$
and $T_l(\beta)$ is the transfer function
for which the same cosmological model as in \cite{Aurich_Lustig_2012c}
is used, see also Section \ref{sec:CMB-correlations}.
The formula (\ref{Eq:Cl_ensemble}) allows to to derive
the minimal parameter range of $(\rho,\alpha,\epsilon)$
for which the whole CMB variability is exhaust.
This variability is exhaust if the quadratic sum of the expansion coefficients
$\xi^{j,i}_{lm}({\cal M};t)$ covers all possible values.
This quadratic sum can be evaluated to
\begin{eqnarray}
\nonumber
\frac{1}{2l+1}\sum_{m=-l}^{l}& &\sum_{i=1}^{r^{\cal M}(\beta)}
\left|\xi^{j,i}_{lm}({\cal M};t)\right|^2 \\
\label{Eq:quadratic_sum_da_TZ_OZ_IZ}
=\frac{1}{2l+1}\sum_{m=-l}^{l}& &\sum_{i=1}^{r^{\cal M}(\beta)}
\left|\sum_{\tilde{m}_b}\langle jm_aj\tilde{m}_b|lm\rangle  \, a^{s(i)}_{m_a} \,
e^{-\hbox{\scriptsize i}\,\tilde{m}_b\,(\alpha-\epsilon)}\,
d^{\,j}_{\tilde{m}_b, m_b(i)}(-2 \rho)\right|^2\\
\nonumber
=\frac{1}{2l+1}\sum_{m,i}& &
\left|{\sum_{m_a}}'\langle jm_aj\,m-m_a|lm\rangle  \, a^{s(i)}_{m_a} \,
e^{\hbox{\scriptsize i}\,m_a\,(\alpha-\epsilon)}\, d^{\,j}_{m-m_a, m_b(i)}(-2 \rho)\right|^2
\hspace{10pt},
\end{eqnarray}
where the prime at the sum over $m_a$ indicates
that the summation is restricted by the
selection rule (\ref{Eq:selection_rule_ma}).
The values of $m_b(i)$ have to be compatible with (\ref{Eq:selection_rule_mb}),
of course.
In the second step of (\ref{Eq:quadratic_sum_da_TZ_OZ_IZ})
the summation over $\tilde{m}_b$ is replaced by $m_a$ using 
$\tilde{m}_b=m-m_a$.
Since only the combination $\alpha-\epsilon$ occurs in the last equation,
one can restrict the CMB analysis to $\alpha=\hbox{const.}$ or
$\epsilon=\hbox{const.}$ and nevertheless screens the whole CMB variability.
In the following we set the coordinate $\epsilon$ to $\epsilon=0$.
Furthermore, the sum is invariant under the substitution
$\alpha \rightarrow  \alpha+2\pi\,k/N$, $k=1,\dots,N-1$, 
because of the selection rule (\ref{Eq:selection_rule_ma}).
This invariance reduces the necessary screening interval of $\alpha$
to $\alpha \in [0,2\pi/N]$.
Since the complete variation of the $d$ function is covered
by the interval $[0,\pi]$,
the complete observer dependence can then be analysed by the
coordinates $\rho\in [0,\pi/2 ]$, $\alpha \in [0,2\pi/N]$.

A further reduction of the $\rho$ interval to $\rho\in [0,\pi/4 ]$
follows from the invariance of the sum due to the transformation 
$(\rho,\alpha)\rightarrow (\pi/2-\rho,\pi/N+\alpha)$.
This invariance can be derived by using the relation 
$d^{\,j}_{m-m_a, m_b}(2 \rho-\pi)=(-1)^{j-2m_b+m_a-m}d^{\,j}_{m-m_a, -m_b}(-2 \rho)$
and by replacing the sum over $m_b$ by a sum over $-m_b$.

An additional invariance is derived in \ref{sec:eigenmodes_polyhedral_da}
which states that the sum (\ref{Eq:quadratic_sum_da_TZ_OZ_IZ}) is invariant
with respect to $(\alpha-\epsilon) \rightarrow -(\alpha-\epsilon)$.
When this invariance is with $\epsilon=0$ rewritten as
$\alpha \rightarrow 2\pi/N-\alpha$
the final screening intervals $\rho\in [0,\pi/4 ]$,
$\alpha \in [0,\pi/N]$, and $\epsilon=0$ are obtained
where all possible ensemble averages for the CMB statistics are encountered.

\section{CMB correlations on large angular scales}
\label{sec:CMB-correlations}

In our previous investigations concerning double-action manifolds
\cite{Aurich_Lustig_2012b,Aurich_Lustig_2012c},
we analysed the CMB statistics in terms of
the temperature 2-point correlation function
\begin{equation}
\label{Eq:C_theta}
C(\vartheta) \; := \; \left< \delta T(\hat n) \delta T(\hat n')\right>
\hspace{10pt} \hbox{with} \hspace{10pt}
\hat n \cdot \hat n' = \cos\vartheta
\hspace{10pt} ,
\end{equation}
where $\delta T(\hat n)$ is the temperature fluctuation in
the direction of the unit vector $\hat n$.
The temperature correlation function $C(\vartheta)$ is computed by
\begin{equation}
\label{Eq:C_theta_C_l}
C(\vartheta) \; = \; \sum_l \frac{2l+1}{4\pi} \, C_l \, P_l(\cos\vartheta)
\end{equation}
using (\ref{Eq:Cl_ensemble}) for the calculation of the multipole moments $C_l$.
From the correlation function $C(\vartheta)$ the scalar statistical measure
\begin{equation}
\label{Eq:S_statistic_60}
S \; := \; \int^{\cos(60^\circ)}_{\cos(180^\circ)}
\hbox{d} \cos\vartheta \; |C(\vartheta)|^2
\end{equation}
is obtained \cite{Spergel_et_al_2003}
which is well suited to measure the suppression of CMB correlations
on angular scales with $\vartheta\gtrsim 60^\circ$.
It has the advantage that it maps the correlation function onto a
scalar quantity which facilitates the comparison of a large number of models.
Since the considered multiconnected spaces are inhomogeneous,
the correlation measure $S$ depends on the observer position defined by the
parameters $(\rho,\alpha)$.
Of special interest is thus the minimum of the $S$ statistics
over the position parameters  $(\rho,\alpha)$
\begin{equation}
\label{Eq:S_min}
S_{\hbox{\scriptsize min}(\alpha,\rho)} \; = \;
\hbox{min}_{\{ \alpha,\rho \}}\left(\frac{S( \alpha,\rho)}{S_{{\cal S}^3}}\right)
\end{equation}
as a function of the total density $\Omega_{\hbox{\scriptsize tot}}$.
The minimum (\ref{Eq:S_min}) is normalised to the
corresponding statistics of the simply connected ${\cal S}^3$ space
in order to emphasise the topological signature.

The correlation measure $S$ has the advantage that it is a property of the
system itself independent of the observed CMB correlations.
However, it is nevertheless important to compare the CMB correlations
of the double-action manifolds with the observed ones.
To that aim the integrated weighted temperature correlation difference
\cite{Aurich_Janzer_Lustig_Steiner_2007}
\begin{equation}
\label{Eq:I_measure}
I := \int_{-1}^1 \hbox{d} \cos\vartheta \; \;
\frac{(C^{\hbox{\scriptsize model}}(\vartheta)-
C^{\hbox{\scriptsize obs}}(\vartheta))^2}
{\hbox{Var}(C^{\hbox{\scriptsize model}}(\vartheta))}
\hspace{10pt} 
\end{equation}
is also analysed,
where the cosmic variance is computed using
$\hbox{Var}(C(\vartheta)) \approx
\sum_l \frac{2l+1}{8\pi^2} \left[C_l \,P_l(\cos\vartheta)\right]^2$.
Similar to the $S$ statistics we also consider the minimum
of the $I$ statistics
\begin{equation}
\label{Eq:I_min}
I_{\hbox{\scriptsize min}(\alpha,\rho)}(\Omega_{\hbox{\scriptsize tot}})
\; = \;
\hbox{min}_{\{ \alpha,\rho\}} \;
I(\alpha,\rho,\Omega_{\hbox{\scriptsize tot}})
\end{equation}
with respect to the model parameters.


\begin{figure}
\hspace*{80pt}\includegraphics[width=9.0cm]{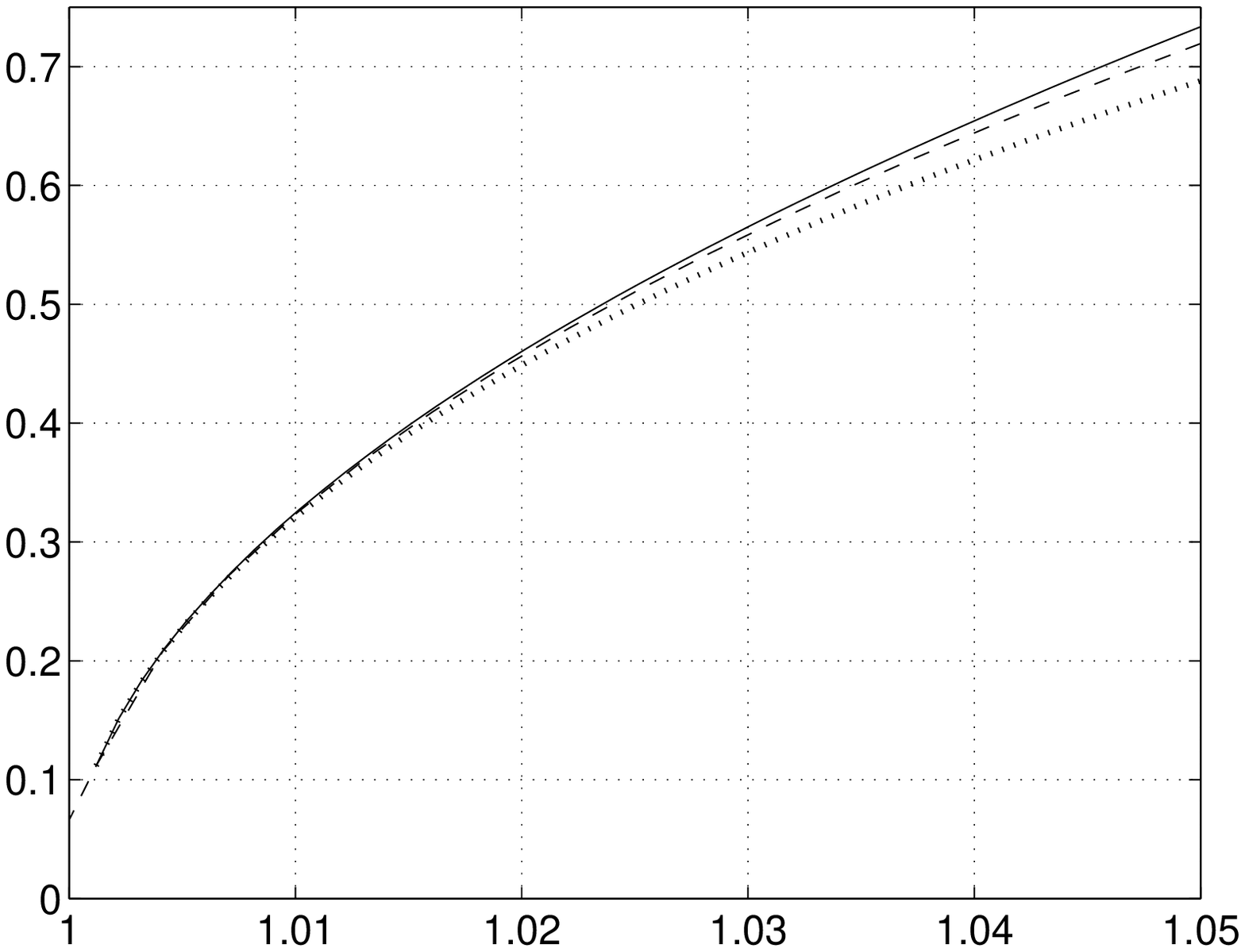}
\put(-255,155){$\tau_{\hbox{\scriptsize sls}}$}
\put(-10,12){$\Omega_{\hbox{\scriptsize tot}}$}
\vspace*{-10pt}
\caption{\label{Fig:tau_sls}
The conformal distance $\tau_{\hbox{\scriptsize sls}}$
to the surface of last scattering is shown in dependence on
$\Omega_{\hbox{\scriptsize tot}}$.
The full curve is obtained by varying only $\Omega_\Lambda$
as it is the case in our simulations.
Alternatively, the variation in $\Omega_{\hbox{\scriptsize tot}}$ is achieved
by changing only $\Omega_{\hbox{\scriptsize mat}}$ in the dotted curve
and by changing the Hubble parameter $h$ in the dashed curve.
}
\end{figure}

The following statistical analysis is based on the same
cosmological parameters as in \cite{Aurich_Lustig_2012c}
which are close to the standard concordance model of cosmology
\cite{Larson_et_al_2011}.
The parameters are taken from the LAMBDA website (lambda.gsfc.nasa.gov),
where we select the WMAP cosmological parameters of the model 'olcdm+sz+lens'
using the data 'wmap7+bao+snconst',
which are $\Omega_{\hbox{\scriptsize b}} = 0.0485$,
$\Omega_{\hbox{\scriptsize cdm}} = 0.238$, the Hubble constant $h=0.681$,
and the spectral index $n_{\hbox{\scriptsize s}}=0.961$.
The total density parameter $\Omega_{\hbox{\scriptsize tot}}$ is varied
by altering the density parameter of the cosmological constant
$\Omega_{\scriptsize \Lambda }$, so that the total density covers the
interval $\Omega_{\hbox{\scriptsize tot}}=1.001,\dots,1.05$.
This $\Omega_{\hbox{\scriptsize tot}}$ interval is a bit larger than the 99\% CL
interval of the constraint $0.99 < \Omega_{\hbox{\scriptsize tot}} < 1.02$
(95\% CL)
which belongs to the chosen set of cosmological parameters.
Our analysis of polyhedral double-action manifolds
covers more than 2.6 million simulations
which are computed for the different values of $\Omega_{\hbox{\scriptsize tot}}$
up to $\Omega_{\hbox{\scriptsize tot}}= 1.05$ and for
different observer positions.
This large number of simulations is the reason
why we restrict our variation of $\Omega_{\hbox{\scriptsize tot}}$
to a variation in $\Omega_{\scriptsize \Lambda}$.
There are other ways of varying $\Omega_{\hbox{\scriptsize tot}}$,
but since the main effect on the CMB on large angular scales is due
to the distance $\tau_{\hbox{\scriptsize sls}}$ to the surface of last scattering,
it suffice to confine to one method of variation.
In order to emphasise this fact,
figure \ref{Fig:tau_sls} shows $\tau_{\hbox{\scriptsize sls}}$ 
as a function of $\Omega_{\hbox{\scriptsize tot}}$
whereas the modification of $\Omega_{\hbox{\scriptsize tot}}$ is achieved
in three different ways,
i.\,e.\ by varying only $\Omega_{\scriptsize \Lambda}$ (full curve),
by varying only $\Omega_{\hbox{\scriptsize mat}}$ (dotted curve)
and by varying the Hubble parameter $h$ (dashed curve).
As seen in figure \ref{Fig:tau_sls}, the three curves differ only
for values of $\Omega_{\hbox{\scriptsize tot}}$ towards
$\Omega_{\hbox{\scriptsize tot}}=1.05$.
Thus, for the analysis of topological suppressions of correlations
on large angles,
the manner in which the change in
$\Omega_{\hbox{\scriptsize tot}}=1.05$ is realized
has only a minor influence on the following results.


\begin{figure}
\vspace*{-15pt}
\hspace*{0.5cm}\begin{minipage}{18.0cm}
\hspace*{-15pt}
\begin{minipage}{9.0cm}
{
\hspace*{-15pt}
\includegraphics[width=9.0cm]{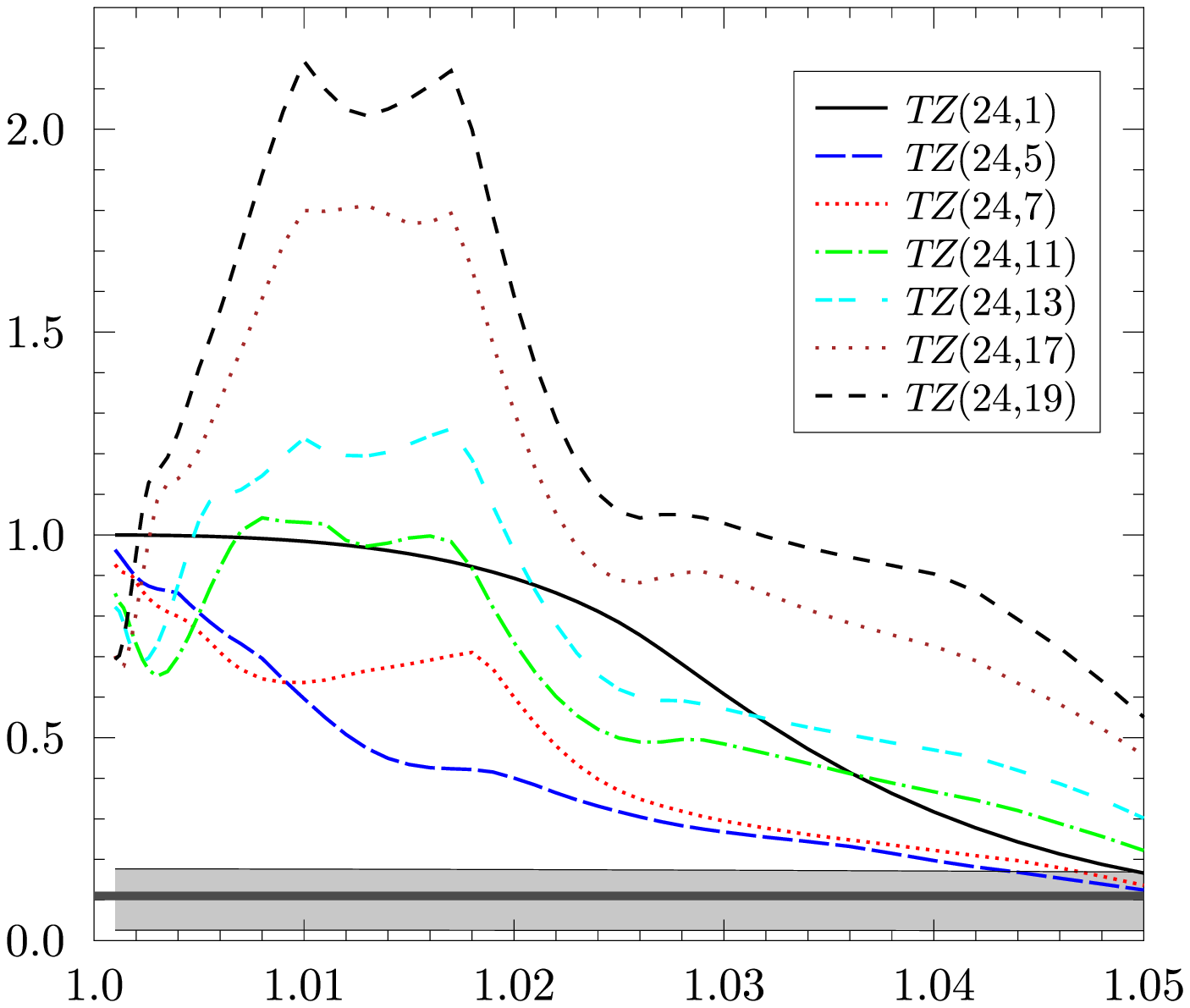}
\put(-250,90){\rotatebox{90}{$S_{\hbox{\scriptsize min}(\alpha,\rho)}$}}
\put(-62,16){$\Omega_{\hbox{\scriptsize tot}}$}
\put(-130,170){(a)}
\vspace*{-45pt}
}
{
\includegraphics[width=9.0cm]{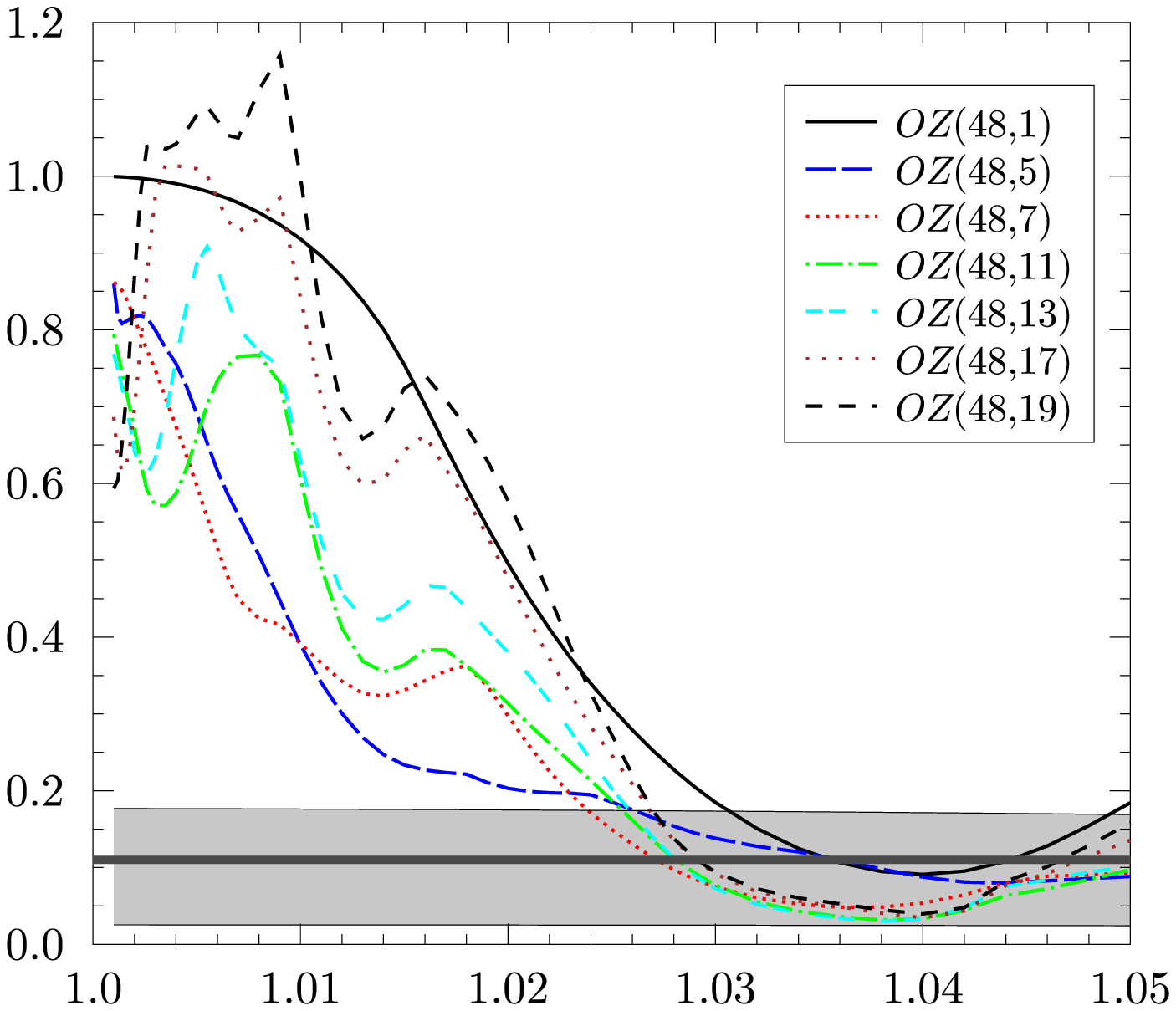}
\put(-250,90){\rotatebox{90}{$S_{\hbox{\scriptsize min}(\alpha,\rho)}$}}
\put(-62,16){$\Omega_{\hbox{\scriptsize tot}}$}
\put(-130,170){(c)}
}
\end{minipage}
\hspace*{-35pt}
\begin{minipage}{9.0cm}
{
\includegraphics[width=9.0cm]{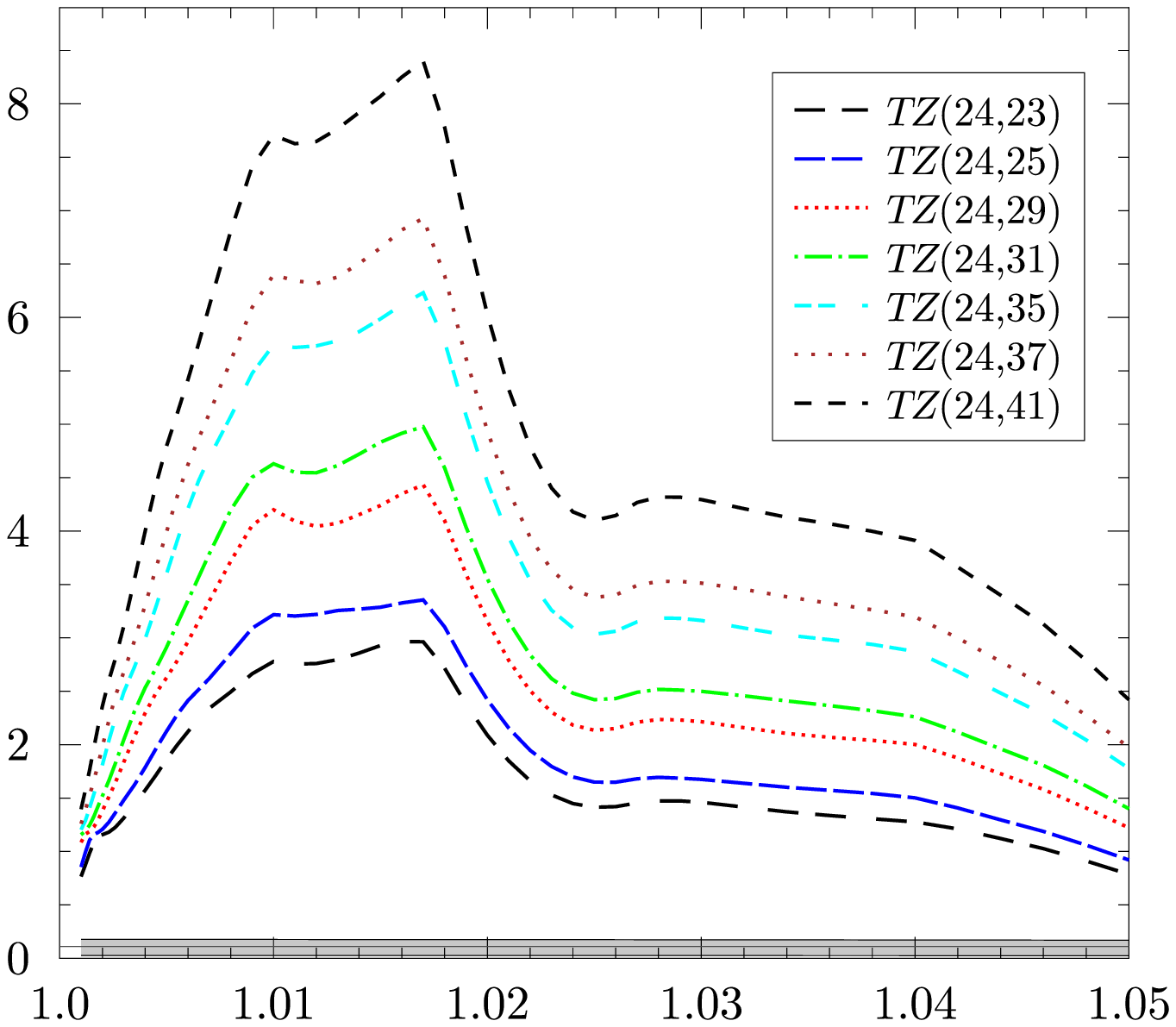}
\put(-250,90){\rotatebox{90}{$S_{\hbox{\scriptsize min}(\alpha,\rho)}$}}
\put(-62,16){$\Omega_{\hbox{\scriptsize tot}}$}
\put(-130,170){(b)}
\vspace*{-45pt}
}
{
\includegraphics[width=9.0cm]{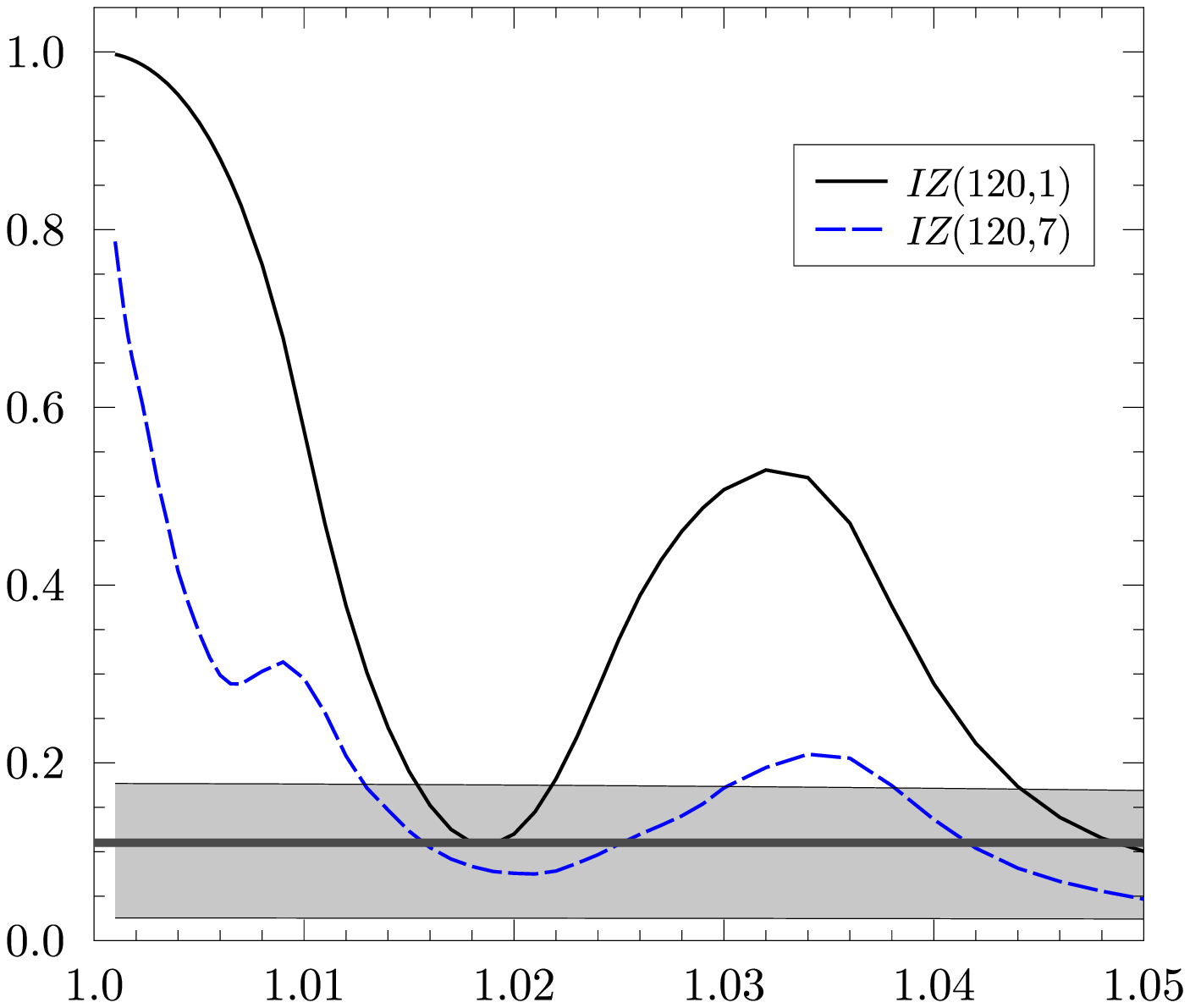}
\put(-250,90){\rotatebox{90}{$S_{\hbox{\scriptsize min}(\alpha,\rho)}$}}
\put(-62,16){$\Omega_{\hbox{\scriptsize tot}}$}
\put(-130,170){(d)}
}
\end{minipage}
\end{minipage}
\caption{\label{Fig:s_statistic_min}
The minima $S_{\hbox{\scriptsize min}(\alpha,\rho)}$ of the $S$ statistics
defined in eq.\,(\ref{Eq:S_min}) are shown for all polyhedral double-action
manifolds up to a group order of 1\,000
as a function of the total density $\Omega_{\hbox{\scriptsize tot}}$.
The horizontal thick line at 0.11 allows the comparison with the
minima of the polyhedral spaces ${\cal T}\equiv TZ(24,1)$,
${\cal O}\equiv OZ(48,1)$, and ${\cal I}\equiv IZ(120,1)$
whose $\Omega_{\hbox{\scriptsize tot}}$-dependent values are shown
as full black curves.
The grey band indicates the range for the $S$ statistics
obtained from the ILC seven year map with and without the KQ75 mask.
}
\end{figure}

The figure \ref{Fig:s_statistic_min} shows
$S_{\hbox{\scriptsize min}(\alpha,\rho)}$ defined in eq.\,(\ref{Eq:S_min})
where the minimum of the $S$ statistics is taken over all observer positions
$(\alpha,\rho)$ possessing distinct CMB ensemble averages.
The four panels show $S_{\hbox{\scriptsize min}(\alpha,\rho)}$ for all
polyhedral double-action manifolds whose group order is below 1\,000.
In order to compare the model results with the observed ones,
the correlation function $C^{\hbox{\scriptsize obs}}(\vartheta)$ is computed
from the ILC 7 year map \cite{Gold_et_al_2010} which gives 
$S_{\hbox{\scriptsize ILC}}(60^\circ) = 8\,033\,\mu\hbox{K}^4$.
By applying the KQ75 7yr mask \cite{Gold_et_al_2010} to the ILC 7 year map,
a correlation function $C^{\hbox{\scriptsize obs}}(\vartheta)$
is obtained which leads to the even lower value
$S_{\hbox{\scriptsize ILC,KQ75}}(60^\circ) = 1\,153\,\mu\hbox{K}^4$.
Both values can be considered as an estimate of the boundaries of the
uncertainty range,
since the KQ75 7yr mask is the most conservative mask and
applying no mask is the other extreme point of view.
The application of the KQ75 7yr mask eliminates the pixels
whose CMB fluctuations are obscured by foreground emissions
mainly originating in the Galaxy.
This range for the observed $S$ statistics in shown in
figure \ref{Fig:s_statistic_min} as the grey horizontal band
where  we have taken our normalisation into account.
Note that the normalisation to the simply connected ${\cal S}^3$ space
gives for the concordance model a value of one.
This emphasises the discrepancy due to this correlation measure.
The analysis of \cite{Copi_Huterer_Schwarz_Starkman_2008} shows
that only 0.025 per cent of realisations of the concordance model
possess such a low correlation.

The 13 double-action manifolds based on the binary tetrahedral space ${\cal T}$
are distributed over the panels (a) and (b) in ascending order.
The correlation measure $S_{\hbox{\scriptsize min}(\alpha,\rho)}$ of the
binary tetrahedral space ${\cal T}\equiv TZ(24,1)$ is shown in panel (a).
Its first minimum occurs at $\Omega_{\hbox{\scriptsize tot}}\simeq 1.07$ and
lies outside the displayed range $\Omega_{\hbox{\scriptsize tot}}\in[1.001,1.05]$.
The minima of the three binary polyhedral spaces ${\cal T}$, ${\cal O}$, and
${\cal I}$ lie close together about $0.11$
which is indicated by the horizontal thick line.
A significantly stronger suppression of CMB correlations than for ${\cal T}$
is revealed by the spaces $TZ(24,5)$ and $TZ(24,7)$.
At the boundary of the 95\% CL interval of $\Omega_{\hbox{\scriptsize tot}}=1.02$,
the best candidate is the $TZ(24,5)$ space with
$S_{\hbox{\scriptsize min}(\alpha,\rho)}\simeq 0.4$.
The space $TZ(24,11)$ behaves approximately as ${\cal T}$.
But for higher group orders $n$ of the cyclic group $Z_n$,
a systematic increase of $S_{\hbox{\scriptsize min}(\alpha,\rho)}$ is observed,
so that these models with larger values of $n$
do not provide viable models for the description of our Universe.
The figure \ref{Fig:s_statistic_min}(b) shows this monotonic increase for the
spaces obtained from $Z_{23}$ up to $Z_{41}$.

For the double-action manifolds derived from the binary octahedral space
${\cal O}$, there are more interesting space forms as revealed by
figure \ref{Fig:s_statistic_min}(c).
The octahedral double-action spaces $OZ(48,5)$ to $OZ(48,17)$ possess an
even stronger suppression for most of the considered values of
$\Omega_{\hbox{\scriptsize tot}}$ compared to ${\cal O}\equiv OZ(48,1)$.
For the space ${\cal O}$ the strongest CMB suppression occurs 
close to $\Omega_{\hbox{\scriptsize tot}}=1.04$.
Several octahedral double-action spaces possess values of
$S_{\hbox{\scriptsize min}(\alpha,\rho)}$ that are even lower than
the best value of $\sim 0.11$ of the three binary polyhedral spaces
${\cal T}$,  ${\cal O}$, and  ${\cal I}$, see the interval
$\Omega_{\hbox{\scriptsize tot}}\in[1.03,1.04]$ in panel (c).
As can be read off from the figure,
the octahedral double-action manifolds $OZ(48,n)$,
with $n=7$, 11, 13, 17, and 19 have suppression factors below 0.11.
The space with $n=5$ obtains its minimum slightly above
$\Omega_{\hbox{\scriptsize tot}}=1.04$.
The table \ref{Tab:double_action_OZ_IZ} lists the positions corresponding to
the minima $S_{\hbox{\scriptsize min}(\Omega_{\hbox{\scriptsize tot}},\alpha,\rho)}$.
Except for $OZ(48,19)$ the minima occur at nearly the same positions
in the $\alpha$-$\rho$ plane.
Furthermore, the correlation measure $S_{\hbox{\scriptsize min}(\alpha,\rho)}$
of the space $OZ(48,19)$ displays a similar behaviour as those of ${\cal O}$
for smaller values of $\Omega_{\hbox{\scriptsize tot}}$.
In contrast to the tetrahedral double-action manifolds $TZ(24,n)$,
there is no simple behaviour with respect to the increase of
$S_{\hbox{\scriptsize min}(\alpha,\rho)}$ in terms of $n$ for the class $OZ(48,n)$
for $n\leq 19$.
For $\Omega_{\hbox{\scriptsize tot}}\leq 1.02$,
the best candidate is $OZ(48,5)$ with
$S_{\hbox{\scriptsize min}(\alpha,\rho)}\simeq 0.2$.
The spaces with $n=7$, 11, and 13 have also a pronounced suppression of
$S_{\hbox{\scriptsize min}(\alpha,\rho)} \simeq 0.30$, 0.31, and 0.38,
respectively, at $\Omega_{\hbox{\scriptsize tot}}=1.02$.


\begin{table}[tbp]
\centering
\begin{tabular}{|c|c|c|c|c|}
\hline
manifold ${\cal M}$ &
$S_{\hbox{\scriptsize min}(\Omega_{\hbox{\scriptsize tot}},\alpha,\rho)}$ &
$\Omega_{\hbox{\scriptsize tot}}$  & $\rho$ &  $\alpha$ \\
\hline
$OZ(48,5)$  & 0.080 & 1.044 & 0.141 & 0.785 \\
$OZ(48,7)$  & 0.048 & 1.036 & 0.134 & 0.785 \\
$OZ(48,11)$ & 0.032 & 1.038 & 0.141 & 0.785 \\
$OZ(48,13)$ & 0.030 & 1.038 & 0.141 & 0.785 \\
$OZ(48,17)$ & 0.035 & 1.040 & 0.157 & 0.785 \\
$OZ(48,19)$ & 0.040 & 1.040 & 0.778 & 0.481 \\
\hline
$IZ(120,7)$ & 0.075 & 1.021 & 0.126 & 0.628 \\
\hline
\end{tabular}
\caption{\label{Tab:double_action_OZ_IZ}
The parameters $ \Omega_{\hbox{\scriptsize tot}},\rho,\alpha$ for which
$S_{\hbox{\scriptsize min}(\Omega_{\hbox{\scriptsize tot}},\alpha,\rho)}$
reveals a local minimum are listed
for the 6 double action manifolds $OZ(24,n)$, $n=5$, 7, 11, 13, 17, and 19
and for the double action manifold $IZ(120,7)$.
}
\end{table}


There exists only one icosahedral double-action space $IZ(120,n)$
whose group order is below 1\,000,
and that is the space $IZ(120,7)$.
Its behaviour is compared to the binary icosahedral space
${\cal I}\equiv IZ(120,1)$ in figure \ref{Fig:s_statistic_min}(d).
It is seen that the suppression is more pronounced for $IZ(120,7)$
than for ${\cal I}$.
As it was the case for some octahedral double-action spaces,
there is again a density range $\Omega_{\hbox{\scriptsize tot}}$
with a suppression stronger than $\sim 0.11$.
The table \ref{Tab:double_action_OZ_IZ} gives the position of the
minimum at $\Omega_{\hbox{\scriptsize tot}}=1.021$
which is slightly larger than the values close to
$\Omega_{\hbox{\scriptsize tot}}=1.05$.
The first minimum of $S_{\hbox{\scriptsize min}(\alpha,\rho)}$
at $\Omega_{\hbox{\scriptsize tot}}=1.007$ has the remarkable
suppression of $S_{\hbox{\scriptsize min}(\alpha,\rho)}=0.27$
which is smaller than the best values of all investigated spherical
manifolds for $\Omega_{\hbox{\scriptsize tot}}\leq 1.01$.

Therefore, among the polyhedral double-action spaces are examples
for multiconnected spaces that display a large suppression
of CMB correlations for angle separations larger
than $\vartheta\geq 60^\circ$.


\begin{figure}
\vspace*{-15pt}
\hspace*{0.5cm}\begin{minipage}{18.0cm}
\hspace*{-15pt}
\begin{minipage}{9.0cm}
{
\hspace*{-15pt}
\includegraphics[width=9.0cm]{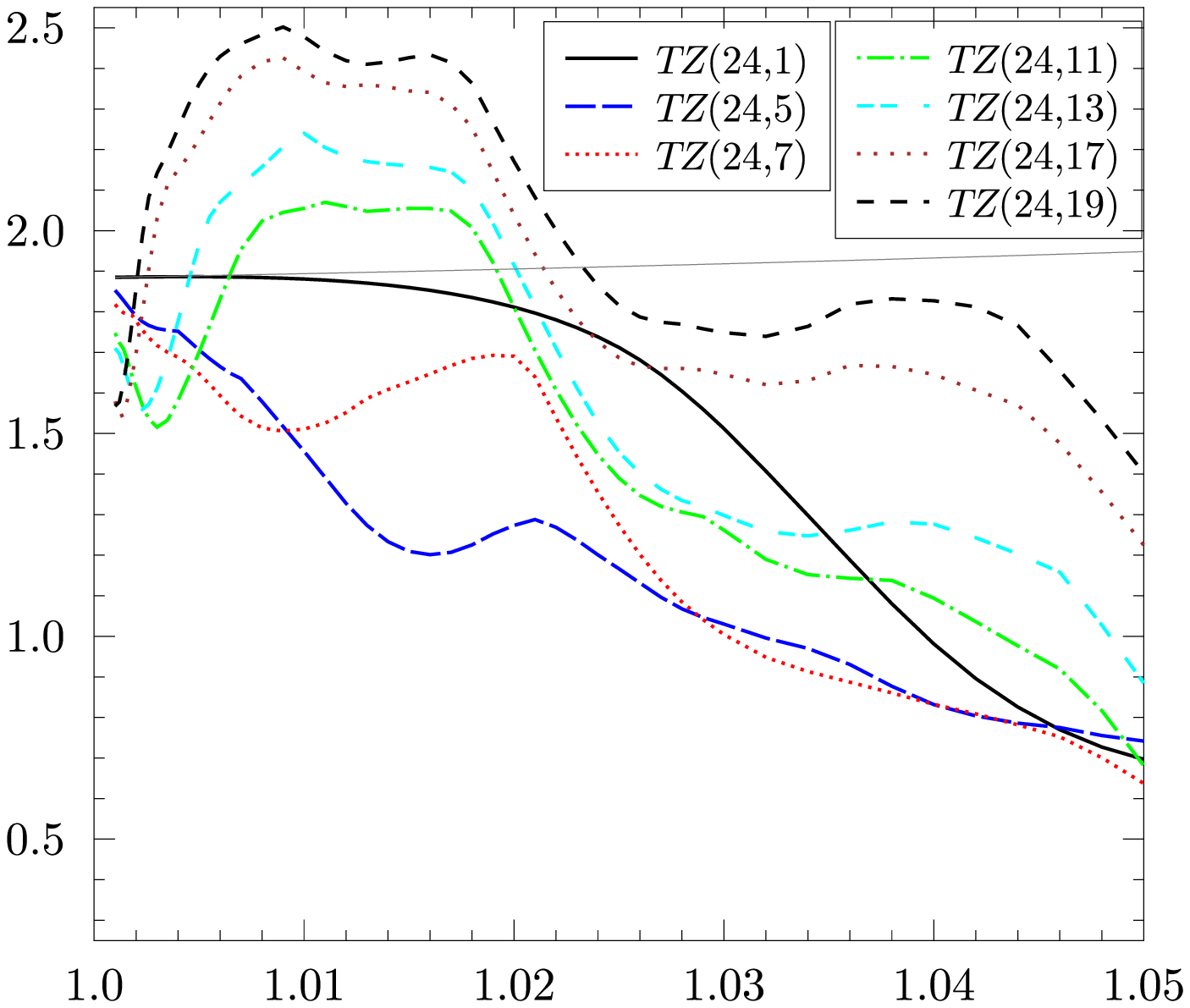}
\put(-250,90){\rotatebox{90}{$I_{\hbox{\scriptsize min}(\alpha,\rho)}$}}
\put(-62,16){$\Omega_{\hbox{\scriptsize tot}}$}
\put(-200,45){(a) no mask}
\vspace*{-45pt}
}
{
\includegraphics[width=9.0cm]{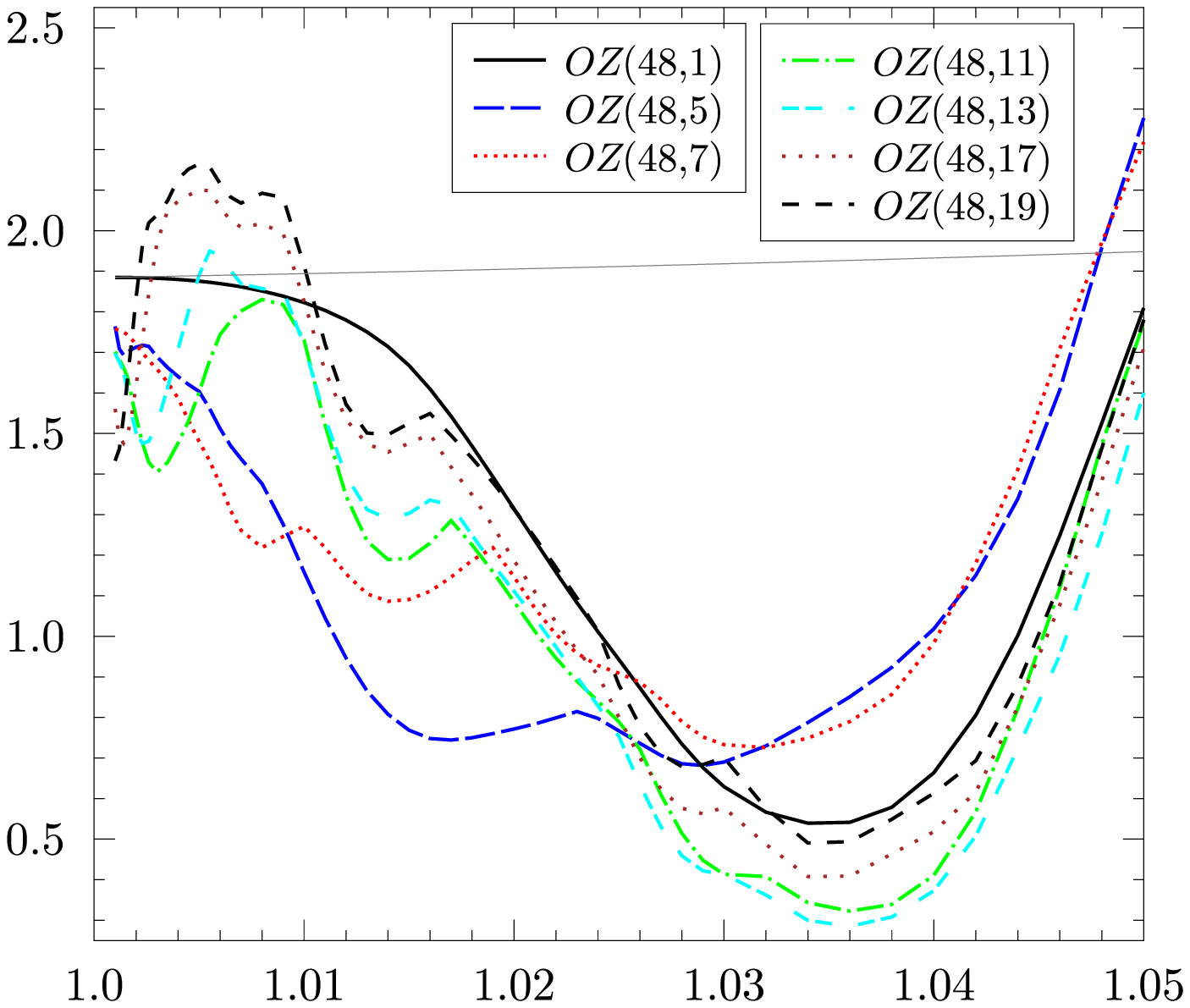}
\put(-250,90){\rotatebox{90}{$I_{\hbox{\scriptsize min}(\alpha,\rho)}$}}
\put(-62,16){$\Omega_{\hbox{\scriptsize tot}}$}
\put(-200,45){(c) no mask}
}
\end{minipage}
\hspace*{-35pt}
\begin{minipage}{9.0cm}
{
\includegraphics[width=9.0cm]{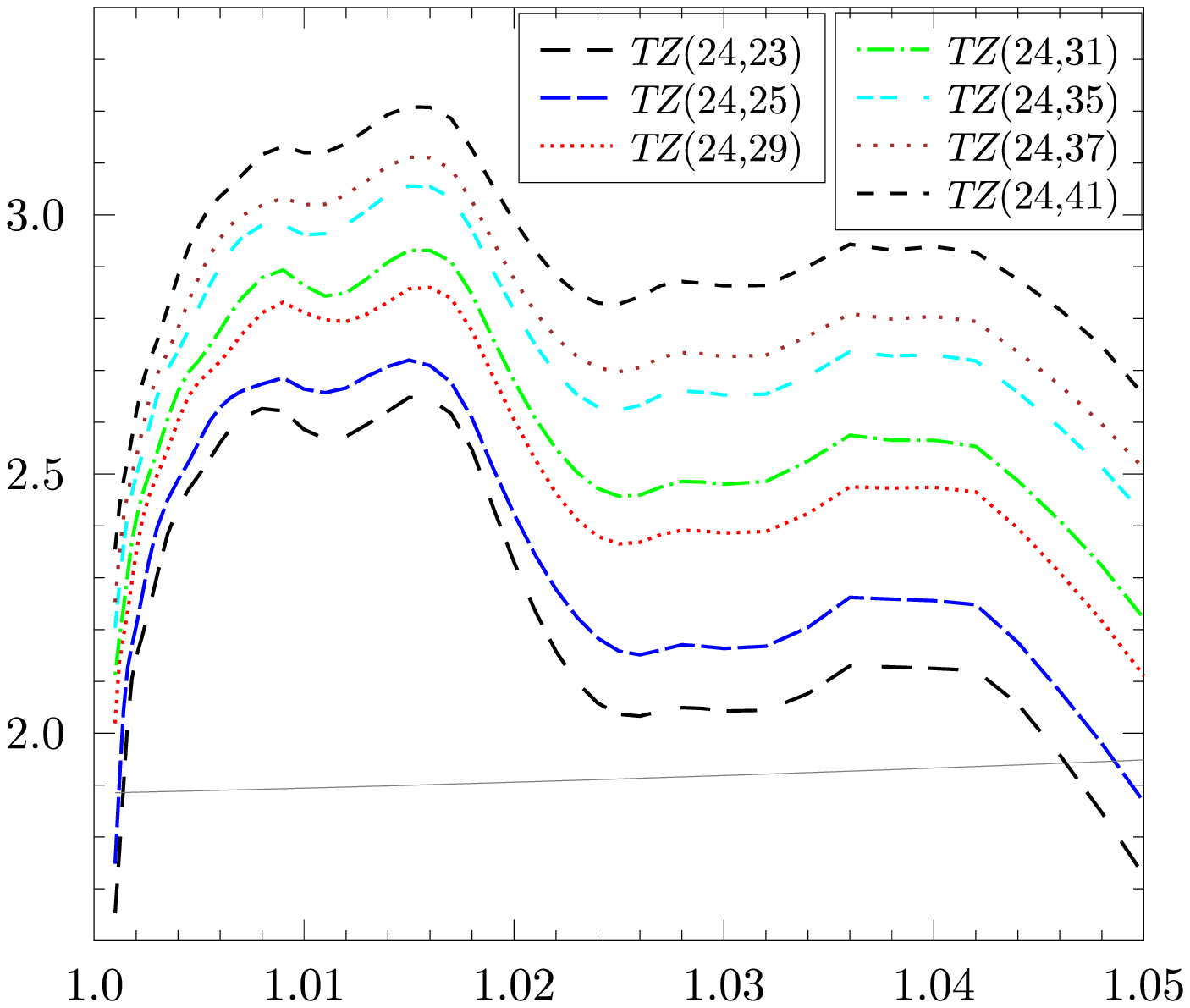}
\put(-250,90){\rotatebox{90}{$I_{\hbox{\scriptsize min}(\alpha,\rho)}$}}
\put(-62,16){$\Omega_{\hbox{\scriptsize tot}}$}
\put(-200,45){(b) no mask}
\vspace*{-45pt}
}
{
\includegraphics[width=9.0cm]{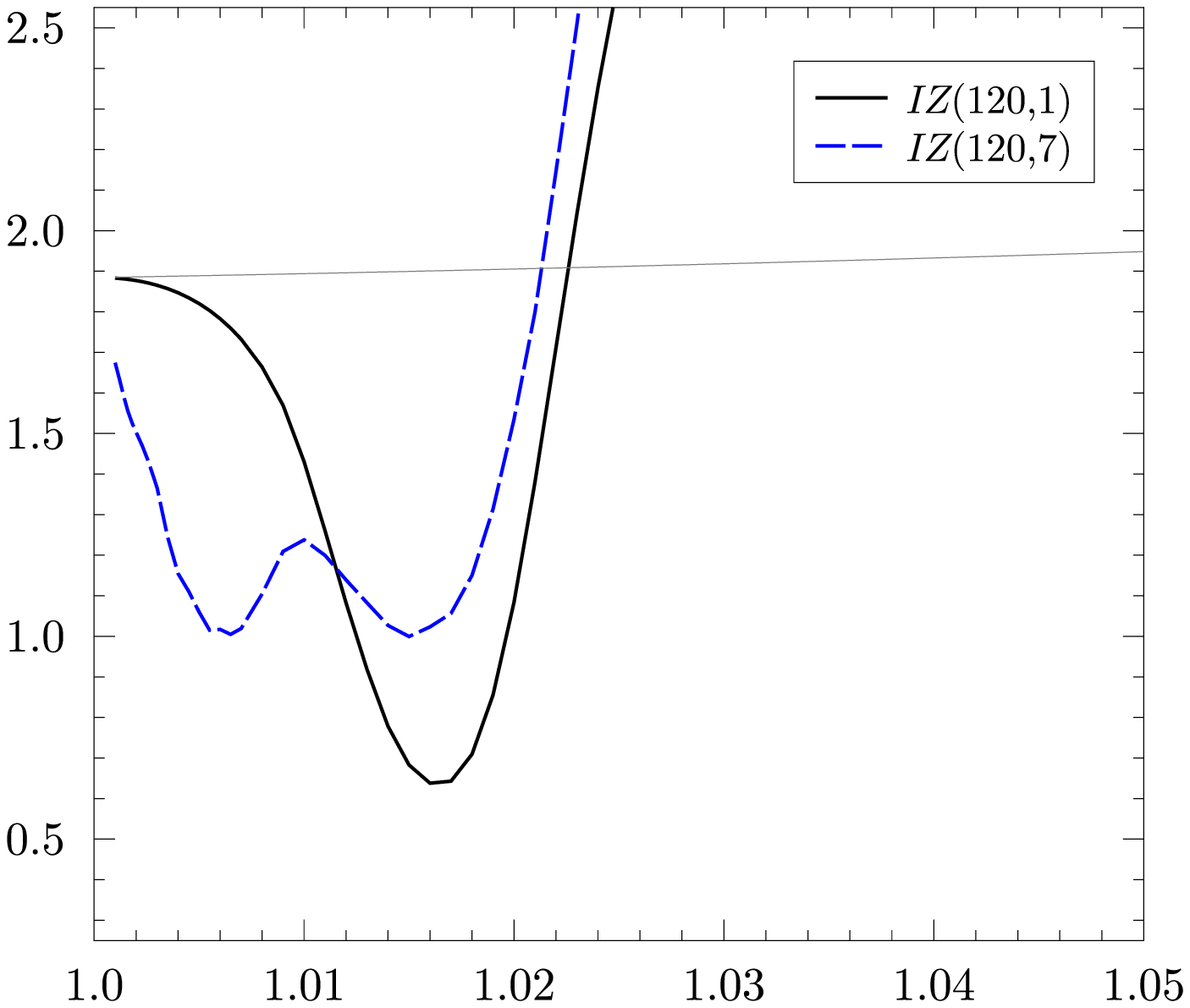}
\put(-250,90){\rotatebox{90}{$I_{\hbox{\scriptsize min}(\alpha,\rho)}$}}
\put(-62,16){$\Omega_{\hbox{\scriptsize tot}}$}
\put(-200,45){(d) no mask}
}
\end{minipage}
\end{minipage}
\caption{\label{Fig:I_statistic_min_no_mask}
The minima
$I_{\hbox{\scriptsize min}(\alpha,\rho)}(\Omega_{\hbox{\scriptsize tot}})$,
defined in eq.\,(\ref{Eq:I_min}), are shown for
the tetrahedral double-action manifolds $TZ(24,n)$ in panels (a) and (b),
the octahedral double-action manifolds $OZ(48,n)$ in panel (c),
and for the icosahedral double-action manifold $IZ(120,7)$ in panel (d).
The double-action correlation functions are compared to the observed
correlation function $C^{\hbox{\scriptsize obs}}(\vartheta)$ obtained from
the WMAP 7 year ILC map without applying any mask.
The full grey curve shows $I(\Omega_{\hbox{\scriptsize tot}})$
for the spherical 3-space ${\cal S}^3$,
i.\,e.\ for the simply connected space.
}
\end{figure}


\begin{figure}
\vspace*{-15pt}
\hspace*{0.5cm}\begin{minipage}{18.0cm}
\hspace*{-15pt}
\begin{minipage}{9.0cm}
{
\hspace*{-15pt}
\includegraphics[width=9.0cm]{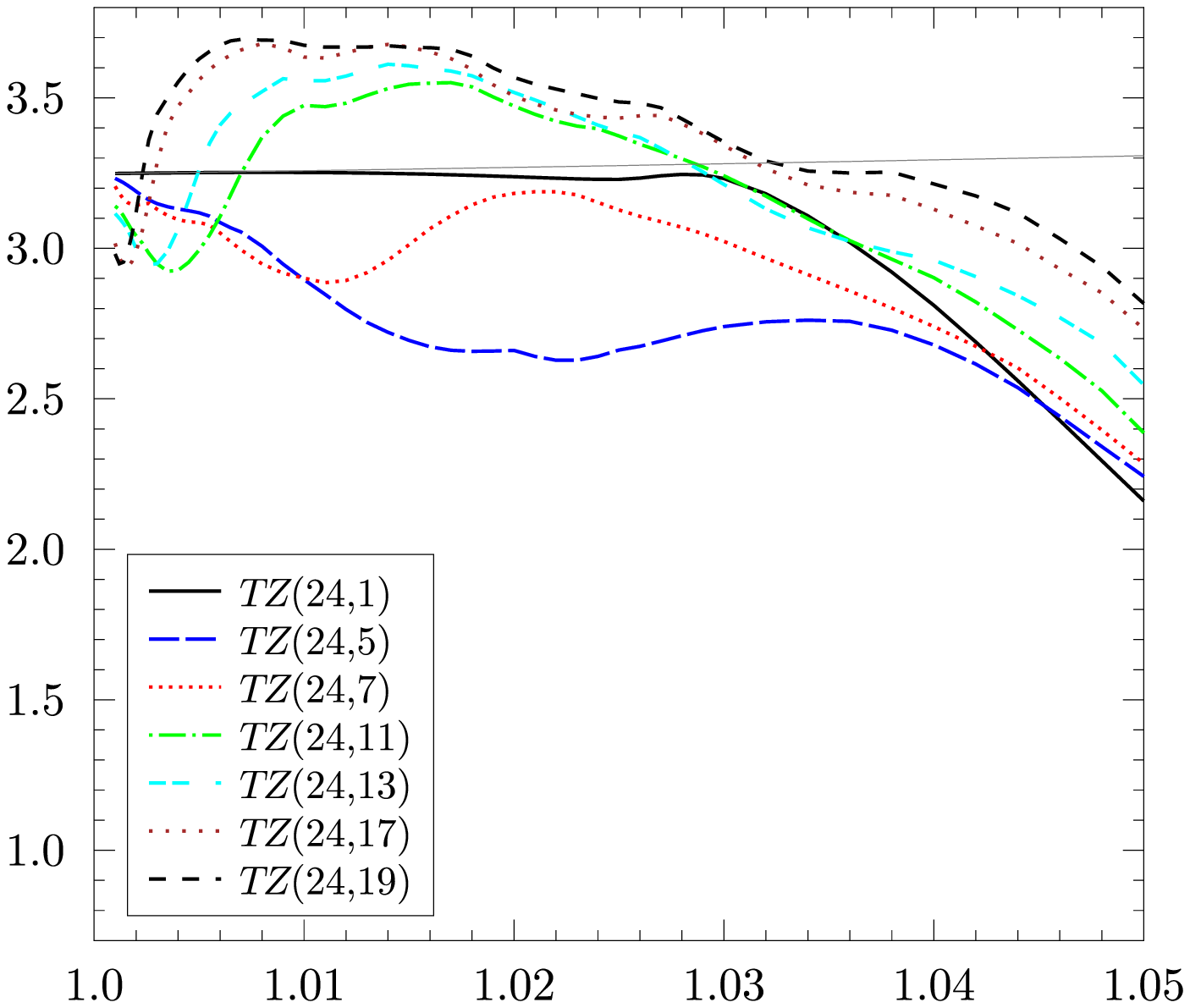}
\put(-250,90){\rotatebox{90}{$I_{\hbox{\scriptsize min}(\alpha,\rho)}$}}
\put(-62,16){$\Omega_{\hbox{\scriptsize tot}}$}
\put(-121,179){(a) KQ75 mask}
\vspace*{-45pt}
}
{
\includegraphics[width=9.0cm]{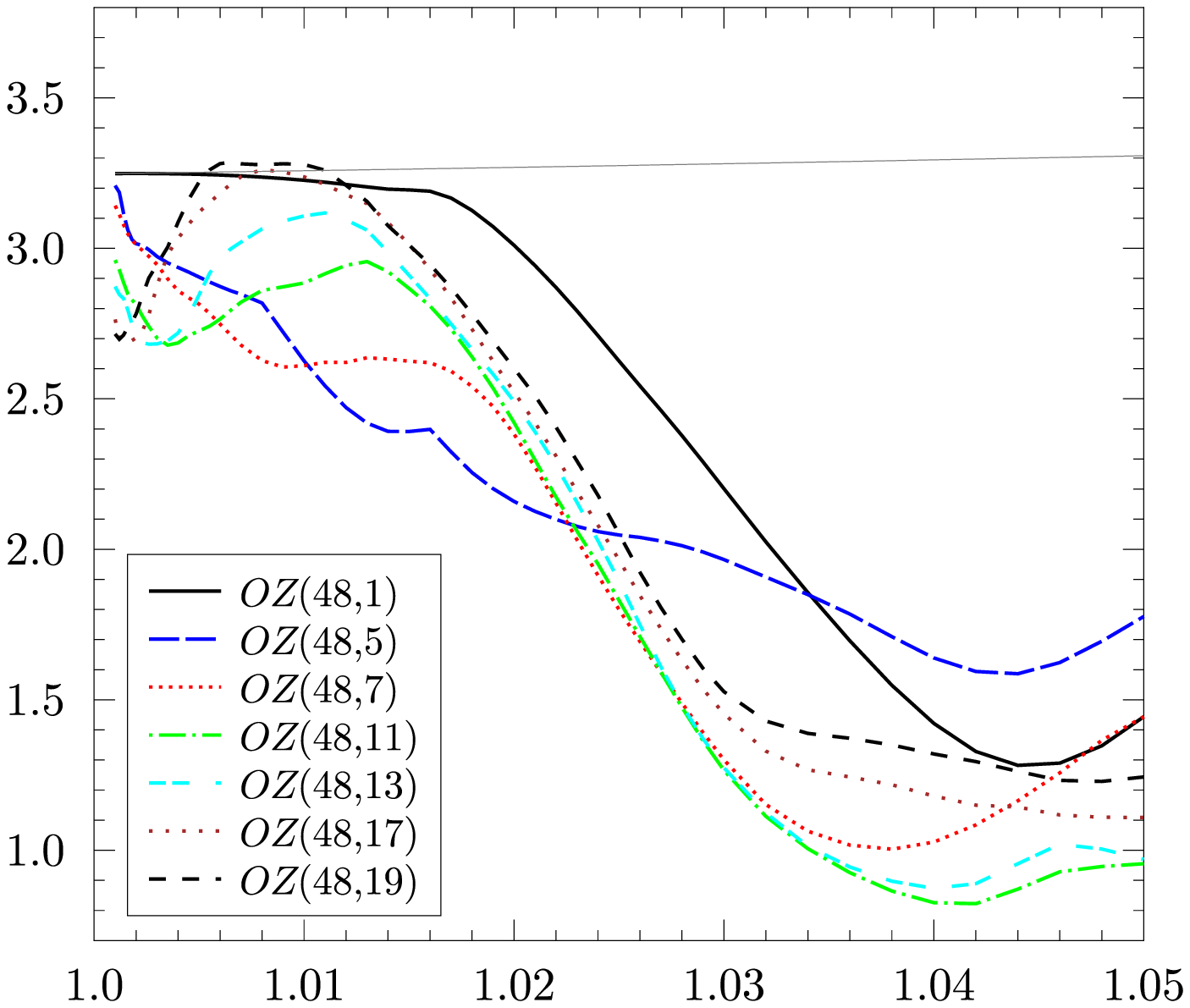}
\put(-250,90){\rotatebox{90}{$I_{\hbox{\scriptsize min}(\alpha,\rho)}$}}
\put(-62,16){$\Omega_{\hbox{\scriptsize tot}}$}
\put(-121,179){(c) KQ75 mask}
}
\end{minipage}
\hspace*{-35pt}
\begin{minipage}{9.0cm}
{
\includegraphics[width=9.0cm]{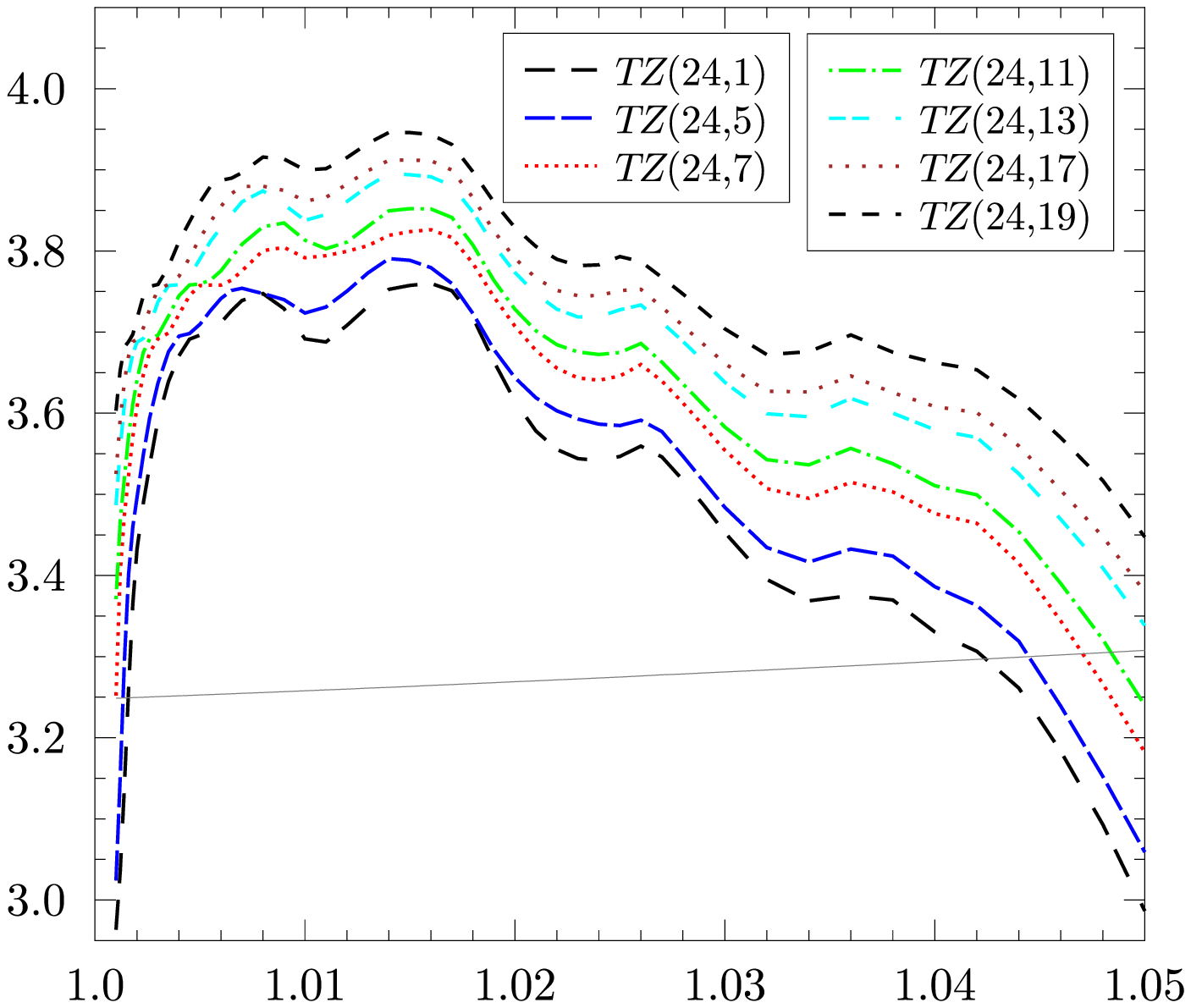}
\put(-250,90){\rotatebox{90}{$I_{\hbox{\scriptsize min}(\alpha,\rho)}$}}
\put(-62,16){$\Omega_{\hbox{\scriptsize tot}}$}
\put(-200,45){(b) KQ75 mask}
\vspace*{-45pt}
}
{
\includegraphics[width=9.0cm]{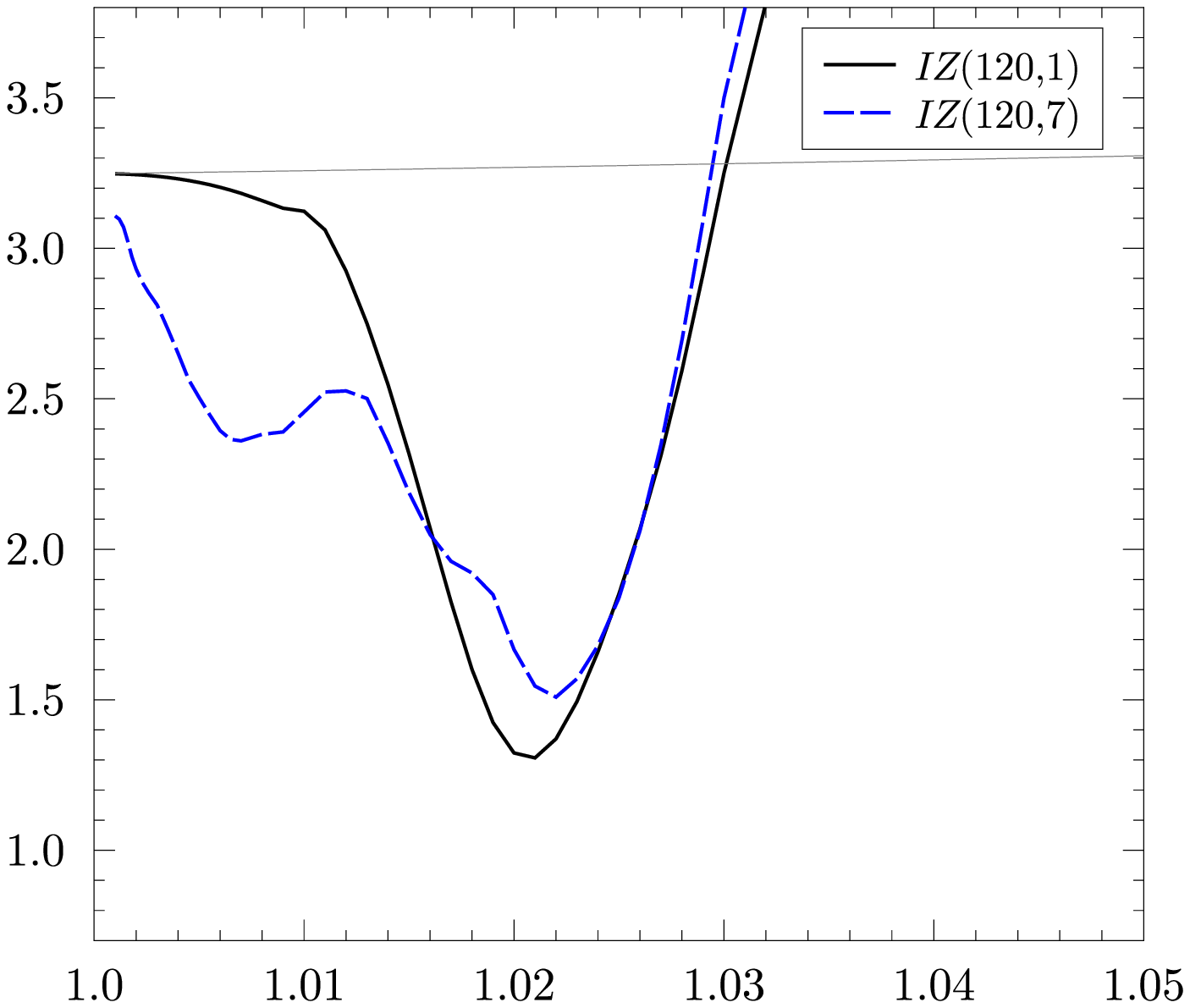}
\put(-250,90){\rotatebox{90}{$I_{\hbox{\scriptsize min}(\alpha,\rho)}$}}
\put(-62,16){$\Omega_{\hbox{\scriptsize tot}}$}
\put(-200,45){(d) KQ75 mask}
}
\end{minipage}
\end{minipage}
\caption{\label{Fig:I_statistic_min_KQ75_mask}
This figure also shows
$I_{\hbox{\scriptsize min}(\alpha,\rho)}(\Omega_{\hbox{\scriptsize tot}})$
as in figure \ref{Fig:I_statistic_min_no_mask},
but now the observed correlation function $C^{\hbox{\scriptsize obs}}(\vartheta)$
is obtained from the WMAP 7 year ILC map by applying the KQ75 mask.
}
\end{figure}

While we have just discussed the correlation measure $S$,
which provides a direct description of the large-angle behaviour of
the multiconnected spaces, we now turn to
the integrated weighted temperature correlation difference $I$,
defined in eq.\,(\ref{Eq:I_measure}).
It reveals how well the ensemble averages
of the correlation functions $C(\vartheta)$ of the double-action spaces
match $C^{\hbox{\scriptsize obs}}(\vartheta)$,
which derives from the observed single realisation of the CMB sky
admissible to us.
The ILC seven year map is used for the computation of the
observed correlation function $C^{\hbox{\scriptsize obs}}(\vartheta)$.
In order to provide an impression of the experimental accuracy,
the analysis is carried out with the full ILC map as well as
with the ILC map subjected to the KQ75 seven year mask.
As discussed above, the differences between these two analyses
reflect the accuracy of the data.
An alternative choice would be to use the $C^{\hbox{\scriptsize obs}}(\vartheta)$
obtained from the W or V band maps,
but it is shown in \cite{Copi_Huterer_Schwarz_Starkman_2008}
that the correlation functions are very similar to those belonging
to the ILC map after applying the KQ75 seven year mask.
Since no significantly changed result is expected,
we restrict us in the following to the ILC map.

The minima $I_{\hbox{\scriptsize min}(\alpha,\rho)}(\Omega_{\hbox{\scriptsize tot}})$
are shown for all polyhedral double-action spaces up to the
group order 1\,000 in figures \ref{Fig:I_statistic_min_no_mask}
and \ref{Fig:I_statistic_min_KQ75_mask} as a function of
$\Omega_{\hbox{\scriptsize tot}}$.
The curves belonging to the multiconnected spaces should be compared
to the simply connected case, i.\,e.\ the spherical 3-space ${\cal S}^3$,
which is shown as the almost horizontal grey curve in
figures \ref{Fig:I_statistic_min_no_mask}
and \ref{Fig:I_statistic_min_KQ75_mask}.
The double-action correlations describe the observed data better than
those of the simply connected space if they lie below the full grey curve.

The tetrahedral double-action manifolds $TZ(24,n)$ are displayed
in panels (a) and (b) of the figures \ref{Fig:I_statistic_min_no_mask}
and \ref{Fig:I_statistic_min_KQ75_mask}.
The general trend for the increasing strength of the correlations
with increasing group order $n$ of the cyclic group $Z_n$,
which was already discovered in the analysis of the $S$ statistics,
is also reflected in the behaviour of
$I_{\hbox{\scriptsize min}(\alpha,\rho)}(\Omega_{\hbox{\scriptsize tot}})$.
The spaces $TZ(24,5)$ and $TZ(24,7)$ give a better match to the
observed data than the binary tetrahedral space ${\cal T}$
which in turn describes the data better than the simply connected space.
Except for values of $\Omega_{\hbox{\scriptsize tot}}$ very close to one,
the models with $n>20$ do not present interesting alternatives.
Note that the quality of the match to the data deteriorates systematically
with increasing values of $n$.
Because of these large values of
$I_{\hbox{\scriptsize min}(\alpha,\rho)}(\Omega_{\hbox{\scriptsize tot}})$,
the panels \ref{Fig:I_statistic_min_no_mask}(b) and
\ref{Fig:I_statistic_min_KQ75_mask}(b) use a different scaling
compared to panels (a), (c), and (d).

The systematic behaviour shown in panels \ref{Fig:I_statistic_min_no_mask}(b)
and \ref{Fig:I_statistic_min_KQ75_mask}(b) is not repeated in the case of the
octahedral double-action manifolds $OZ(48,n)$ which are displayed in panel (c).
For $\Omega_{\hbox{\scriptsize tot}}$ below
$\Omega_{\hbox{\scriptsize tot}} \simeq 1.025$
there is a sequence of $OZ(48,n)$ spaces which provides the best description
of the data.
With decreasing value of $\Omega_{\hbox{\scriptsize tot}}$,
these are the spaces with $n=5$, $7$, and $11$,
see panels \ref{Fig:I_statistic_min_no_mask}(c)
and \ref{Fig:I_statistic_min_KQ75_mask}(c).
For values of $\Omega_{\hbox{\scriptsize tot}}$ larger than $1.025$,
however, one finds in the case without a mask four spaces with smaller
values of $I_{\hbox{\scriptsize min}(\alpha,\rho)}(\Omega_{\hbox{\scriptsize tot}})$
which even beats the minimum of the binary octahedral space ${\cal O}$.
These are the spaces $OZ(48,13)$, $OZ(48,11)$, $OZ(48,17)$, and $OZ(48,19)$.
Applying the KQ75 mask to the ILC data,
also the curve belonging to the $OZ(48,7)$ space drops below
that of the binary octahedral space ${\cal O}$.

The icosahedral double-action manifold $IZ(120,7)$ does not lead to
a better agreement with the data than the binary icosahedral space ${\cal I}$
at that value of $\Omega_{\hbox{\scriptsize tot}}$
where the latter space has its minimum in
$I_{\hbox{\scriptsize min}(\alpha,\rho)}(\Omega_{\hbox{\scriptsize tot}})$.
But for smaller values of $\Omega_{\hbox{\scriptsize tot}}$,
the space $IZ(120,7)$ describes the data better than ${\cal I}$
as can be seen in figures \ref{Fig:I_statistic_min_no_mask}(d)
and \ref{Fig:I_statistic_min_KQ75_mask}(d).

The figures \ref{Fig:I_statistic_min_no_mask}
and \ref{Fig:I_statistic_min_KQ75_mask} bring out the quality of the
description of the data with respect to the data of the full ILC map
as well as to the data restricted by the KQ75 mask.
The comparison between figure \ref{Fig:I_statistic_min_no_mask}
and figure \ref{Fig:I_statistic_min_KQ75_mask} reveals
that the polyhedral double-action manifolds give a better match to
the correlation function $C^{\hbox{\scriptsize obs}}(\vartheta)$
derived from the full ILC map.
Furthermore, the positions of the minima of
$I_{\hbox{\scriptsize min}(\alpha,\rho)}(\Omega_{\hbox{\scriptsize tot}})$
are shifted to larger values of $\Omega_{\hbox{\scriptsize tot}}$
when the KQ75 mask is applied.
This behaviour is, for example, visible in the panels
\ref{Fig:I_statistic_min_no_mask}(d) and \ref{Fig:I_statistic_min_KQ75_mask}(d)
where the binary icosahedral space ${\cal I}$ possesses a minimum
at $\Omega_{\hbox{\scriptsize tot}} \simeq 1.016$ without mask
and at $\Omega_{\hbox{\scriptsize tot}} \simeq 1.021$ with KQ75 mask.
This demonstrates that the choice of the available data leads
to a range of variation so that only general properties of the
double-action manifolds can be inferred from figures
\ref{Fig:I_statistic_min_no_mask} and \ref{Fig:I_statistic_min_KQ75_mask}.


\begin{table}[tbp]
\centering
\begin{tabular}{|c||c|c||c|c|}
\hline
manifold ${\cal M}$ &
$I_{\hbox{\scriptsize min}(\Omega_{\hbox{\scriptsize  
tot}},\alpha,\rho)}^{\hbox{\scriptsize no mask}}$ &
$\Omega_{\hbox{\scriptsize tot}}$  & $I_{\hbox{\scriptsize  
min}(\Omega_{\hbox{\scriptsize tot}},\alpha,\rho)}^{\hbox{\scriptsize   
KQ75 mask}}$ &  $\Omega_{\hbox{\scriptsize tot}}$ \\
\hline
${\cal S}^3$ & 1.885 & 1.001 & 3.249 & 1.001 \\
\hline
$OZ(48,5)$  & 0.744 & 1.017 & 2.159 & 1.020 \\
$OZ(48,7)$  & 1.086 & 1.014 & 2.383 & 1.020 \\
$OZ(48,11)$ & 1.085 & 1.020 & 2.421 & 1.020 \\
$OZ(48,13)$ & 1.112 & 1.020 & 2.491 & 1.020 \\
$OZ(48,17)$ & 1.187 & 1.020 & 2.523 & 1.020 \\
$OZ(48,19)$ & 1.313 & 1.020 & 2.601 & 1.020 \\
\hline
$IZ(120,7)$ & 0.999 & 1.015 & 1.667 & 1.020 \\
\hline
\end{tabular}
\caption{\label{Tab:I_min_double_action_OZ_IZ}
The table lists the manifolds with the best agreement with the observed
correlation function $C^{\hbox{\scriptsize obs}}(\vartheta)$
which is obtained either from the full ILC map (no mask, columns 2 and 3)
or after applying the KQ75 mask (columns 4 and 5).
The interval of $\Omega_{\hbox{\scriptsize tot}}$ is restricted to
$\Omega_{\hbox{\scriptsize tot}}\le 1.02$.
The value of ${\cal S}^3$, which corresponds to the concordance model,
is also given.
}
\end{table}


Summarising, table \ref{Tab:I_min_double_action_OZ_IZ} gives
the promising models,
which have below $\Omega_{\hbox{\scriptsize tot}} = 1.02$ the most
pronounced minima in the $I$ statistics.
The columns 2 and 3 refer to the analysis without a mask
which is shown in figure \ref{Fig:I_statistic_min_no_mask},
whereas columns 4 and 5 gives the values for the KQ75 mask case
shown in figure \ref{Fig:I_statistic_min_KQ75_mask}.
With the restriction $\Omega_{\hbox{\scriptsize tot}} \leq 1.02$,
the best model is given by $OZ(48,5)$ at $\Omega_{\hbox{\scriptsize tot}}=1.017$,
if the full ILC map is used.
The next best space is provided by $IZ(120,7)$ at
$\Omega_{\hbox{\scriptsize tot}}=1.015$.
Their values of
$I_{\hbox{\scriptsize min}(\Omega_{\hbox{\scriptsize  tot}},\alpha,\rho)}^
{\hbox{\scriptsize no mask}}$
are significantly lower than the value $1.885$
belonging to the concordance model.
Table \ref{Tab:I_min_double_action_OZ_IZ} reveals
that the application of the KQ75 mask leads to minimal values
for the $I$ statistics at the interval boundary
$\Omega_{\hbox{\scriptsize tot}}=1.02$.
The best model is now $IZ(120,7)$ followed by the 
octahedral double-action manifolds $OZ(48,n)$
whose ranking with respect to the $I$ statistics is identical to
the sequence of $n$,
i.\,e.\ they are ranked by their volume.


\begin{figure}
\vspace*{-180pt}\hspace*{60pt}\includegraphics[width=14.0cm]{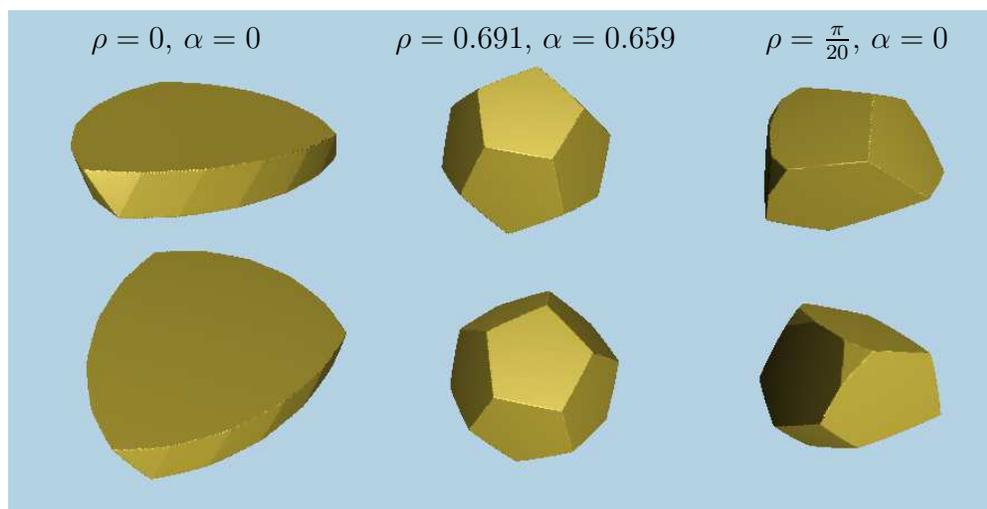}
\put(-355,370){$\rho =0$, $\alpha = 0$}
\put(-240,370){$\rho =0.691$, $\alpha = 0.659$}
\put(-100,370){$\rho =\frac\pi{20}$, $\alpha = 0$}
\vspace*{-185pt}
\caption{\label{Fig:TZ_24_5}
The Dirichlet domain of the tetrahedral double-action manifold $TZ(24,5)$
is shown as seen from three different observer positions.
Two different projections are depicted for each observer position.
At left the observer is at $\rho =0$ and $\alpha = 0$,
in the middle column the position is chosen to be at
$\rho =0.691$ and $\alpha = 0.659$
which corresponds to the shape of the dodecahedron.
The right column shows the Dirichlet domain where the first minimum
occurs in $S_{\hbox{\scriptsize min}(\alpha,\rho)}$ at
$\Omega_{\hbox{\scriptsize tot}} \simeq 1.15$.
}
\end{figure}

In \cite{Aurich_Lustig_2012c} we pointed out
that the two prism double-action manifolds $DZ(8,3)$ and $DZ(16,3)$
possess for a special observer position a Dirichlet domain identical to
the binary tetrahedral space ${\cal T}$ and
to the binary octahedral space ${\cal O}$, respectively.
The Dirichlet domain of the binary icosahedral space ${\cal I}$ does not
emerge among the class of prism double-action manifolds.
This Dirichlet domain, however, is obtained from the
tetrahedral double-action manifold $TZ(24,5)$
again for a special observer position.
In figure \ref{Fig:TZ_24_5} the Dirichlet domains of $TZ(24,5)$ are shown
for three observer positions.
At $\rho =0.691$ and $\alpha = 0.659$ the dodecahedron emerges
which is also the Dirichlet domain of the binary icosahedral space ${\cal I}$.
Thus, the special Dirichlet domains of all three binary polyhedral spaces
can be found within the class of the double-action manifolds.
Also shown is the Dirichlet domain for that observer position
where the largest suppression of CMB correlations on large scales occurs
as measured by $S_{\hbox{\scriptsize min}(\alpha,\rho)}$.
This minimum, which is obtained at $\Omega_{\hbox{\scriptsize tot}} \simeq 1.15$,
corresponds to the Dirichlet domain shown at the right hand side of
figure \ref{Fig:TZ_24_5}.
Remarkably, it is not the most regular Dirichlet domain
which thus demonstrates that oddly shaped domains can lead to a stronger
CMB suppression than well-proportioned ones.

\begin{figure}
\vspace*{0pt}\hspace*{30pt}\includegraphics[width=14.0cm]{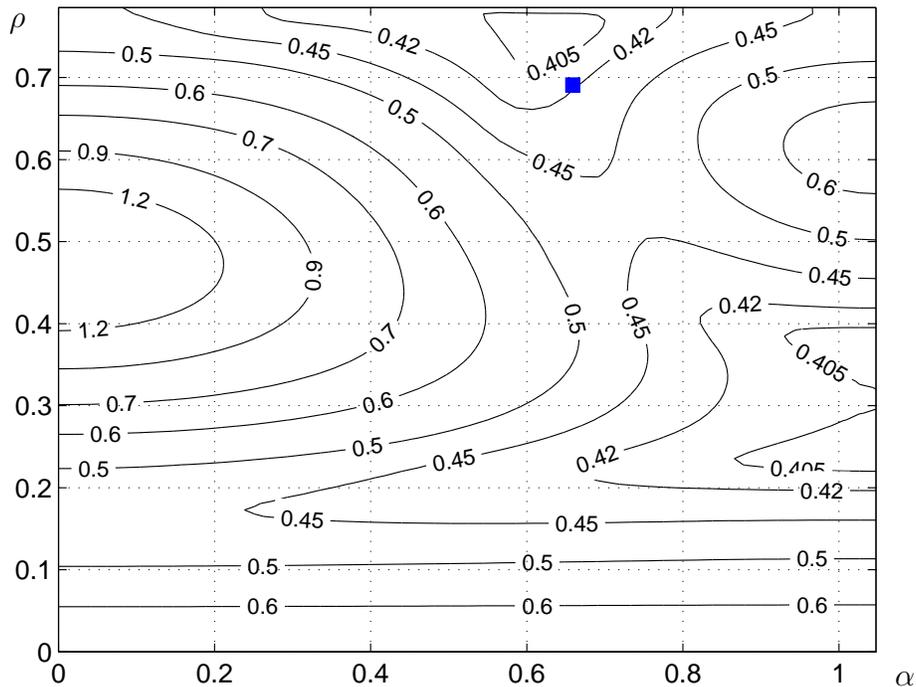}
\put(-365,270){$\rho$}
\put(-30,21){$\alpha$}
\vspace*{-20pt}
\caption{\label{Fig:S_statistics_alpha_rho_dTZ_24_5}
The $\alpha$-$\rho$ dependence of the $S$ statistics is displayed
for the manifold $TZ(24,5)$ at $\Omega_{\hbox{\scriptsize tot}}=1.02$
normalised by the value $S_{{\cal S}^3}$ of the simply connected space.
The full square indicates the position $\rho =0.691$ and $\alpha = 0.659$
where the Dirichlet domain of the space $TZ(24,5)$ has the shape of
the dodecahedron.
}
\end{figure}

This point is emphasised by figure \ref{Fig:S_statistics_alpha_rho_dTZ_24_5}
where the correlation measure $S$ is plotted for
$\Omega_{\hbox{\scriptsize tot}}=1.02$
in such a way that the full observer dependence can be inferred.
The value of $\Omega_{\hbox{\scriptsize tot}}=1.02$ is selected
because at that value the binary icosahedral space ${\cal I}$
provides the best description of the CMB correlations.
The figure reveals a region in the $\alpha$-$\rho$ plane
where the correlation measure $S$ yields values larger than those of
the simply connected ${\cal S}^3$ space.
But besides this region around $\alpha=0$ and $\rho=0.5$,
the values of $S$ drop to values as low as $0.4$.
The minimal values are obtained for three positions
at $(\alpha,\rho)\simeq (0.63,0.74)$,
$(\alpha,\rho)\simeq (\pi/3,0.24)$, and
$(\alpha,\rho)\simeq (\pi/3,0.38)$.
Although the position $\alpha = 0.659$ and $\rho =0.691$
with the dodecahedral Dirichlet domain
is not very far from one of the three minima,
it is nevertheless not the position giving the minimum.

\section{Summary and Discussion}

This paper analyses the large-scale correlations in the
CMB sky for the polyhedral double-action manifolds.
With this analysis, the CMB correlations are finally investigated for all
double-action manifolds
since those belonging to the lens spaces and to the prism double-action
manifolds are already studied in \cite{Aurich_Lustig_2012b} and
\cite{Aurich_Lustig_2012c}.
The large-scale correlation measure (\ref{Eq:S_min}) is used in order
the search for spaces with a significant suppression of
correlations in the CMB anisotropy on scales above $\vartheta>60^\circ$.
This quantity is normalised to the simply connected spherical
space ${\cal S}^3$.
The lens spaces $L(p,q)$ can lead to a suppression relative to ${\cal S}^3$ by
a factor of about $\sim 0.5$ \cite{Aurich_Lustig_2012b}.
The lens spaces with such a large suppression have lenticular fundamental cells
whose two faces have to be rotated by a relative angle of $\sim 101^\circ$
or $\sim 137^\circ$ before the faces are identified.
Among the prism double-action manifolds $DZ(p,n)$,
there are spaces with even smaller large-scale correlations with
suppression factors in the range $0.3\dots 0.4$.
The three best candidates are $DZ(8,3)$,  $DZ(16,3)$, and $DZ(20,3)$
\cite{Aurich_Lustig_2012c}.
Although this CMB suppression is remarkable,
it is less pronounced than in the cases of the
three binary polyhedral spaces ${\cal T}$, ${\cal O}$, and ${\cal I}$
where the suppression factor is of the order of $0.11$.

The three binary polyhedral spaces ${\cal T}$, ${\cal O}$, and ${\cal I}$
lead to the three classes $TZ(24,n)$, $OZ(48,n)$, and $IZ(120,n)$
of polyhedral double-action manifolds.
The analysis of this paper shows
that several polyhedral double-action manifolds can possess
even stronger suppressions than
those found in the three binary polyhedral spaces
(see figure \ref{Fig:s_statistic_min}).
From these spaces, the octahedral double-action manifolds $OZ(48,n)$
with $n=7$, 11, 13, 17, and 19 have suppression factors below 0.11
for $\Omega_{\hbox{\scriptsize tot}}$ in the range
$\Omega_{\hbox{\scriptsize tot}}=1.03\dots 1.04$.
With the constraint $\Omega_{\hbox{\scriptsize tot}}\leq 1.02$,
the best octahedral double-action manifold is the space $OZ(48,5)$
with a suppression factor 0.2.
In addition, three further octahedral spaces with $n=7$, 11, and 13 possess
suppression factors between 0.3 and 0.4
in that $\Omega_{\hbox{\scriptsize tot}}$ range.

The icosahedral double-action manifold $IZ(120,7)$ also reveals
an interesting behaviour with a suppression factor below 0.11
close to $\Omega_{\hbox{\scriptsize tot}}=1.02$.
Remarkably, insisting on the constraint $\Omega_{\hbox{\scriptsize tot}}\leq 1.01$,
the space $IZ(120,7)$ has the largest suppression of all investigated
spherical manifolds.
The minimum in the correlation measure is obtained at
$\Omega_{\hbox{\scriptsize tot}}=1.007$ with a suppression factor of 0.27.

The tetrahedral double-action manifolds $TZ(24,n)$ do not provide
comparable candidates to explain the low correlations on the CMB sky
at large angles.
They possess such small suppression factors only for significantly
larger values of the total density $\Omega_{\hbox{\scriptsize tot}}$
which are beyond the range considered in this paper.
Some $TZ(24,n)$ spaces have nevertheless CMB suppressions comparable
to the prism double-action manifolds $DZ(p,n)$ mentioned above.

The ensemble averages of the correlation functions $C(\vartheta)$
of the polyhedral double-action manifolds are also compared to the
observed $C^{\hbox{\scriptsize obs}}(\vartheta)$
using the $I$ statistics.
This analysis confirms the result obtained from the $S$ statistics.

Concluding, there are five octahedral double-action manifolds
and one icosahedral double-action manifold with a group order below 1\,000
with a pronounced suppression of CMB correlation on large angular scales
which deserve further investigations.

\appendix

\section{Matrix Representations of  $\hbox{SU}(2,\mathbb{C})$}
\label{sec:ap_SU2_matrices}

Every matrix $u \in \hbox{SU}(2,\mathbb{C})$ can be written as
\begin{equation}
\label{su2_matrix_cartesian}
u\; = \; {\bf 1} w + \hbox{i}\, \sigma_x x + \hbox{i}\, \sigma_y y 
                   + \hbox{i}\, \sigma_z z
 \; = \;
 \left( \begin{array}{cc}
w+\hbox{i}\,z & 
\hbox{i}\,\left(x-\hbox{i}\,y\right)\\
\hbox{i}\,\left(x+\hbox{i}\,y\right) &
w-\hbox{i}\,z
\end{array} \right) 
\hspace{10pt}
\end{equation}
with the restriction $w^2+x^2+y^2+z^2=1$.
Here the Pauli matrices are denoted by
$$
\sigma_x\; = \;
 \left( \begin{array}{cc}
0 & 1\\
1& 0
\end{array} \right)\hspace{10pt},\hspace{10pt}
\sigma_y\; = \;
 \left( \begin{array}{cc}
0 & -\hbox{i}\\
\hbox{i}& 0
\end{array} \right)\hspace{10pt},\hspace{10pt} \hbox{and} \hspace{10pt}
\sigma_z\; = \;
 \left( \begin{array}{cc}
1 & 0\\
0& -1
\end{array} \right)
$$
and ${\bf 1}$ is the identity matrix.
Since $\hbox{SU}(2,\mathbb{C})$ can be identified with the
3-sphere ${\cal S}^3$,
the matrix $u$ can be interpreted as a coordinate matrix
which describes points on the 3-sphere ${\cal S}^3$ with the
Cartesian coordinates $(w,x,y,z)$.
An alternative choice of coordinates $(\rho,\alpha,\epsilon)$
on ${\cal S}^3$ is related to these Cartesian coordinates $(w,x,y,z)$ by 
$$
\left( \begin{array}{c}
w\\
x\\
y\\
z
\end{array} \right)
=\left( \begin{array}{c}
\cos\rho\cos\alpha\\
\sin\rho\sin\epsilon\\
\sin\rho\cos\epsilon\\
\cos\rho\sin\alpha
\end{array} \right)
\hspace{10pt}
$$
with $\rho\in[0,\pi/2]$, $\alpha,\epsilon\in[0,2\pi]$.
In terms of the coordinates $(\rho,\alpha,\epsilon)$,
the matrix $u$ reads
\begin{equation}
\label{Eq:su2_matrices_rho_alpha_epsilon}
u(\rho,\alpha,\epsilon) \; = \;
\left( \begin{array}{cc}
\cos\rho\, e^{+\hbox{\scriptsize i}\alpha} &
\sin\rho\, e^{+\hbox{\scriptsize i}\epsilon} \\
-\sin\rho\, e^{-\hbox{\scriptsize i}\epsilon} &
\cos\rho\, e^{-\hbox{\scriptsize i}\alpha}
\end{array} \right)
\hspace{10pt} .
\end{equation}
This parametrisation of a $\hbox{SU}(2,\mathbb{C})$ matrix is used for the
transformation $t$ in section\,\ref{sec:eigenmode},
see eq.\,(\ref{Eq:coordinate_t_rho_alpha_epsilon}).
It facilitates computations involving the Wigner polynomials
since the matrix elements of $u(\rho,\alpha,\epsilon)$ are given
by $D^{1/2}_{\pm1/2,\pm1/2}(u)$.

However, for the analysis involving the spherical coordinates
$(\tau,\theta,\phi$), it is more convenient to use
$$
\left( \begin{array}{c}
w\\
x\\
y\\
z
\end{array} \right)
=\left( \begin{array}{c}
\cos\tau\\
\sin\tau\sin\theta\cos\phi\\
\sin\tau\sin\theta\sin\phi\\
\sin\tau\cos\theta
\end{array} \right)
\hspace{10pt}
$$
with $\tau,\theta\in[0,\pi]$, $\phi\in[0,2\pi]$. 
Then, the matrix $u=u(\tau,\theta,\phi)$ reads 
\begin{equation}
\label{Eq:su2_matrices_spherical_coordinates}
u(\tau,\theta,\phi) \; = \;
\left( \begin{array}{cc}
\cos\tau + \hbox{i}\,\sin\tau\cos\theta &
\hbox{i}\,\sin\tau\sin\theta\,\e^{-\hbox{\scriptsize i}\phi}\\
\hbox{i}\,\sin\tau\sin\theta\,\e^{\hbox{\scriptsize i}\phi}&
\cos\tau - \hbox{i}\,\sin\tau\cos\theta
\end{array} \right)
\hspace{10pt}.
\end{equation}
The origin of the coordinate system $(w,x,y,z)=(1,0,0,0)\equiv{\bf 1}$
can be shifted to the point $u(\tau,\theta,\phi)$ using the transformation
$g(\tau,\theta,\phi):{\bf 1}\rightarrow u(\tau,\theta,\phi)
=g_a^{-1}(\tau,\theta,\phi)\,{\bf 1}\,g_b(\tau,\theta,\phi)$,
where
\begin{equation}
\label{Eq:trafo_spherical_coordinates}
g(\tau,\theta,\phi)\;=\;\left(g_a(\tau,\theta,\phi),g_b(\tau,\theta,\phi)\right)
\end{equation}
with
$$
g_a(\tau,\theta,\phi) \;=\;
 \left(\begin{array}{cc}
 e^{-\hbox{\scriptsize i}\frac{\tau}2}& 0 \\
0 & e^{\hbox{\scriptsize i}\frac{\tau}2}
\end{array} \right)
\left( \begin{array}{cc}
\sin\frac{\theta}2& \cos\frac{\theta}2 \\
-\cos\frac{\theta}2  & \sin\frac{\theta}2 
\end{array} \right)
\left(\begin{array}{cc}
 e^{\hbox{\scriptsize i}\frac{\phi}2}& 0 \\
0 & e^{-\hbox{\scriptsize i}\frac{\phi}2}
\end{array} \right)
$$
and
$$
g_b(\tau,\theta,\phi) \;=\;
 \left(\begin{array}{cc}
 e^{\hbox{\scriptsize i}\frac{\tau}2}& 0 \\
0 & e^{-\hbox{\scriptsize i}\frac{\tau}2}
\end{array} \right)
\left( \begin{array}{cc}
\sin\frac{\theta}2& \cos\frac{\theta}2 \\
-\cos\frac{\theta}2  & \sin\frac{\theta}2 
\end{array} \right)
\left(\begin{array}{cc}
 e^{\hbox{\scriptsize i}\frac{\phi}2}& 0 \\
0 & e^{-\hbox{\scriptsize i}\frac{\phi}2}
\end{array} \right)
\hspace{10pt} .
$$

\section{Eigenmodes on the 3-Sphere in the Spherical Coordinates}
\label{sec:ap_eigenmodes_S3}

The aim of this section is to derive the factorisation of
the eigenmodes of the Laplace-Beltrami operator in terms of
the usual spherical harmonics $Y_{lm}(\theta,\phi)$ and a radial function,
and furthermore, to find a Fourier expansion for the radial function
\cite{Gundermann_2005},
which simplifies the numerical computation of the eigenmodes of
multiconnected spaces.

The eigenmodes are given in spherical coordinates as 
\begin{equation}
\label{Eq:eigenmode_S3_spherical_abstract}
\Psi_{jlm}(\tau,\theta,\phi):=\langle \tau,\theta,\phi|  j; l, m \rangle
= \langle 0,0,0|D(\tau,\theta,\phi)|j; l, m \rangle
\hspace{10pt} 
\end{equation}
with $D(\tau,\theta,\phi)=e^{\hbox{\scriptsize i} \tau \left(-J_{az}+J_{bz}\right)}
\,e^{\hbox{\scriptsize i} \theta (J_{ay}+J_{by})}
\,e^{\hbox{\scriptsize i} \phi \left(J_{az}+J_{bz}\right)}$.
In spherical coordinates, $D(\tau,\theta,\phi)$ shifts the origin $(0,0,0)$
to the point $(\tau,\theta,\phi)$,
where the matrix representation of this transformation 
is given by eq.\,(\ref{Eq:trafo_spherical_coordinates}).
Using the completeness relation
$\sum_{l'm'}|j; l', m' \rangle \langle j; l', m'|={\bf 1}$,
one can rewrite
\begin{equation}
\label{Eq:eigenmode_S3_spherical_umformung_Psi}
\nonumber
\Psi_{jlm}(\tau,\theta,\phi) \; = \;
\sum_{l'm'} \Psi_{jl'm'}(0,0,0) \,
\langle j; l', m'|D(\tau,\theta,\phi)|j; l, m \rangle
\hspace{10pt} ,
\end{equation}
where $\Psi_{jl'm'}(0,0,0)=\langle 0,0,0|j; l', m' \rangle$.
Because of the factorisation $\Psi_{jlm}(\tau,\theta,\phi)\sim
R_{\beta}^{\,l}(\tau)Y_{lm}(\theta,\phi)$ with $\beta=2j+1$, $l=0,\dots,\beta-1$,
one gets the property $\Psi_{jlm}(0,0,0)\sim \delta_{l,0}\delta_{m,0}$
from the spherical harmonics $Y_{lm}(\theta,\phi)$
and from $R_{\beta}^{\,l}(\tau)$ as defined in Eq.\,(A.21)
in \cite{Abbott_Schaefer_1986}.
This simplifies (\ref{Eq:eigenmode_S3_spherical_umformung_Psi}) to
\begin{equation}
\label{Eq:eigenmode_S3_spherical_umformung_no_sum}
\nonumber
\Psi_{jlm}(\tau,\theta,\phi) \; = \;
\Psi_{j00}(0,0,0) \,
\langle j; 0, 0|D(\tau,\theta,\phi)|j; l, m \rangle
\hspace{10pt} .
\end{equation}
With the help of the completeness relation
$\sum_{m'_am'_b}|j; m'_a, m'_b \rangle \langle j; m'_a, m'_b|={\bf 1}$
and the eigenvalue equation
$\langle j; m'_a, m'_b|
e^{\hbox{\scriptsize i} \tau \left(-J_{az}+J_{bz}\right)}=
\langle j; m'_a, m'_b|
e^{\hbox{\scriptsize i} \tau \left(-m'_a+m'_b\right)}$,
the matrix element is manipulated
\begin{eqnarray}
\nonumber
\langle j; l', m'|D(\tau,\theta,\phi)|j; l, m \rangle
\hspace*{-110pt} & &
\\ \nonumber
& = &
\sum_{m'_am'_b} \langle j; l', m'|j; m'_a, m'_b \rangle \langle j; m'_a, m'_b|
D(\tau,\theta,\phi)|j; l, m \rangle
\\ \nonumber
& = &
\sum_{m'_am'_b}e^{\hbox{\scriptsize i} \tau \left(-m'_a+m'_b\right)}
\langle j; l', m'|j; m'_a, m'_b \rangle \langle j; m'_a, m'_b|
D(0,\theta,\phi)|j; l, m \rangle
\\ \label{Eq:D_matrix_lm}
& = &
\sum_{m'_am'_bm''_am''_b}e^{\hbox{\scriptsize i} \tau \left(-m'_a+m'_b\right)}
\langle j; m'_a, m'_b| D(0,\theta,\phi)|j; m''_a, m''_b \rangle
\\ \nonumber
&  & \hspace*{100pt} \times \;
\langle j; l', m'|j; m'_a, m'_b \rangle \,
\langle j; m''_a, m''_b|j; l, m \rangle
\hspace{10pt} .
\end{eqnarray}
Because of the product relation (\ref{Eq:SO4_Basis}), one gets
\begin{eqnarray}
\nonumber
\langle j; m'_a, m'_b|D(0,\theta,\phi)|j; m''_a, m''_b \rangle
\hspace{-110pt} & \; &
\\ \nonumber & = &
\langle j m'_a |e^{\hbox{\scriptsize i} \theta J_{ay}}\,
e^{\hbox{\scriptsize i} \phi J_{az}}|j m''_a\rangle
\langle j m'_b |e^{\hbox{\scriptsize i} \theta J_{by}}\,
e^{\hbox{\scriptsize i} \phi J_{bz}}|j m''_b\rangle
\\ \nonumber & = &
D^{\,j}_{m'_a,m''_a}(0,\theta,\phi)D^{\,j}_{m'_b,m''_b}(0,\theta,\phi)
\\ \nonumber & = &
\sum_{l''}\langle j; m'_a, m'_b| j;l'', m'\rangle 
\langle j;l'', m'' | j; m''_a, m''_b\rangle D^{\,l''}_{m',m''}(0,\theta,\phi)
\hspace{10pt} ,
\end{eqnarray}
where the eq.\,(4.3.1) in \cite{Edmonds_1964} has been used.
Here, the abbreviations $m':= m'_a + m'_b$ and
$m'':= m''_a + m''_b$ have been introduced.
Inserting this result into (\ref{Eq:D_matrix_lm})
leads with
$
\sum_{m''_am''_b}\langle j; l'', m'' | j; m''_a, m''_b\rangle
\langle j; m''_a, m''_b| j; l, m\rangle
= \delta_{l'',l} \delta_{m'',m}
$
to
\begin{eqnarray}
\label{Eq:D_matrix_lm_reduced}
\langle j; l', m'|D(\tau,\theta,\phi)|j; l, m \rangle
\hspace*{-130pt} & &
\\ \nonumber
& = &
\sum_{m'_am'_b}e^{\hbox{\scriptsize i} \tau \left(-m'_a+m'_b\right)}
\langle j; l', m'|j; m'_a, m'_b \rangle \,
\langle j; m'_a, m'_b| j;l, m'\rangle 
D^{\,l}_{m',m}(0,\theta,\phi)
\hspace{10pt} .
\end{eqnarray}
Using the simplified matrix element (\ref{Eq:D_matrix_lm_reduced}),
the eigenmode (\ref{Eq:eigenmode_S3_spherical_umformung_no_sum})
is expressed as
\begin{eqnarray}
\label{Eq:eigenmode_S3_spherical_product}
\nonumber
\Psi_{jlm}(\tau,\theta,\phi)
\nonumber
&=& \Psi_{j00}(0,0,0) \, D^{\,l}_{0,m}(0,\theta,\phi)
\\
\nonumber
& &
\times
\sum_{m'_a}e^{-\hbox{\scriptsize i} \tau 2m'_a}
\langle j; 0, 0|j; m'_a, -m'_a \rangle \,
\langle j; m'_a, -m'_a| j;l, 0\rangle
\hspace{10pt} .
\end{eqnarray}
With the relation
$D^{\,l}_{0,m}(0,\theta,\phi)=\sqrt{\frac{4\pi}{2l+1}}Y_{lm}(\theta,\phi)$,
Eq.\,(4.1.25) in \cite{Edmonds_1964},
the normalisation of the eigenmode
$\Psi_{jlm}(\tau,\theta,\phi)=\tilde{R}_{\beta}^{\,l}(\tau)Y_{lm}(\theta,\phi)$
to ${\cal S}^3$ leads to
$\Psi_{j00}(0,0,0)=\sqrt{\frac{(2j+1)^2}{2\pi^2}}$
which is chosen to be real.
Then the Fourier expansion of the radial function $\tilde{R}_{\beta}^{\,l}(\tau)$
can be read off as
\begin{equation}
\label{Eq:Radial_Function_tilde_R}
\tilde{R}_{\beta}^{\,l}(\tau) \, = \, \sqrt{\frac{2\,(2j+1)^2}{\pi\,(2l+1)}}
\sum_{m'_a}\langle j m'_a j -m'_a| 0 0\rangle \langle j m'_a j -m'_a| l 0\rangle 
e^{-\hbox{\scriptsize i} 2\,\tau m'_a}
\hspace{5pt} ,
\end{equation}
where $\beta=2j+1$ and the notation of the Clebsch-Gordan coefficients
have been introduced.
The radial function $\tilde{R}_{\beta}^{\,l}(\tau)$ differs from
$R_{\beta}^{\,l}(\tau)$ used in \cite{Abbott_Schaefer_1986} by
a phase factor according to
$\tilde{R}_{\beta}^{\,l}(\tau)=(-\hbox{i})^l\,R_{\beta}^{\,l}(\tau)$.

\section{Eigenmodes on Binary Polyhedral Spaces}
\label{sec:eigenmodes_polyhedral}

The eigenmodes of the binary polyhedral spaces 
are already investigated in
\cite{Lachieze_Rey_2004,Lachieze-Rey_Caillerie_2005,Gundermann_2005}.
These eigenmodes can also be
computed as the special case $n=1$ of the ansatz (\ref{Eq:Ansatz_red_Basis}),
which incorporates the action of the generator of the cyclic group $Z_n$
and the action of the first generator (\ref{Def:poly_group})
of the binary  polyhedral groups $T^\star$, $O^\star$, or $I^\star$.
These deck groups lead to the spaces
${\cal M}=TZ(24,1)$, $OZ(48,1)$, and $OZ(120,1)$. 

At first, the ansatz (\ref{Eq:Ansatz_red_Basis}) is expressed in terms of
the spherical coordinates $(\tau,\theta,\phi)$.
Using the completeness relation
$\sum_{lm}|j; l, m \rangle \langle j; l, m|={\bf 1}$
one gets with eq.\,(\ref{Eq:eigenmode_S3_spherical_abstract})
\begin{eqnarray}
\label{Eq:eigenmode_polyhedral_spherical}
\nonumber
\psi^{\, \cal M}_{j;s,m_b}(\tau,\theta,\phi)&:=&\langle \tau,\theta,\phi| j; s, m_b \rangle\\
\nonumber
&=&\sum_{m_a\equiv 0 \;\hbox{\scriptsize mod}\;N}a^s_{m_a}\sum_{l}\langle j m_a j m_b| l m\rangle \Psi_{jlm}(\tau,\theta,\phi)
\hspace{10pt} 
\end{eqnarray}
with $|m_b| \le j$, $N=3$ (binary tetrahedral space),
$N=4$ (binary octahedral space), or $N=5$ (binary icosahedral space).
In the next step, the eigenmodes
$\Psi_{jlm}(\tau,\theta,\phi)=\tilde{R}_{\beta}^{\,l}(\tau)Y_{lm}(\theta,\phi)$
are expressed by the Fourier expansion (\ref{Eq:Radial_Function_tilde_R}).
To constrain the coefficients $a^s_{m_a}$,
the invariance of the eigenmodes under the action of the second generator of
the binary polyhedral group is taken into account by requiring
$\psi^{\, \cal M}_{j;s,m_b}(\tau,\theta_2,\phi_2)-
\psi^{\, \cal M}_{j;s,m_b}(\tau+\tau_2,\theta_2,\phi_2)=0$,
which leads to
\begin{eqnarray}
\label{Eq:equation_polyhedral_1}
\nonumber
& &\sum_{m_a\equiv 0 \;\hbox{\scriptsize mod}\;N}a^s_{m_a}\sum_{lm'_a}\langle j m_a j m_b| l m\rangle \sqrt{\frac{2\,(2j+1)^2}{\pi\,(2l+1)}}
\langle j m'_a j -m'_a| 0 0\rangle \\
\nonumber
& &\hspace{60pt} \times \langle j m'_a j -m'_a| l 0\rangle 
(1-e^{-\hbox{\scriptsize i} 2\,\tau_2 m'_a})e^{-\hbox{\scriptsize i} 2\,\tau m'_a}
Y_{lm}(\theta_2,\phi_2)=0
\hspace{10pt} .
\end{eqnarray}
This equation written as
$\sum_{m_a'}A_{m_a'}e^{-\hbox{\scriptsize i} 2\,\tau m'_a} =0$
has to be valid for every value of $\tau$.
Therefore, each coefficient $A_{m'_a}$ with $m'_a=-j,\dots, j$
has to vanish identically, and one gets $2j+1$ equations
\begin{eqnarray}
\label{Eq:equation_polyhedral_final}
& &\sum_{m_a\equiv 0 \;\hbox{\scriptsize mod}\;N}a^s_{m_a}\sum_{l}\langle j m_a j m_b| l m\rangle \sqrt{\frac{2\,(2j+1)^2}{\pi\,(2l+1)}}
\langle j m'_a j -m'_a| 0 0\rangle \\
\nonumber
& &\hspace{60pt} \times \langle j m'_a j -m'_a| l 0\rangle 
(1-e^{-\hbox{\scriptsize i} 2\,\tau_2 m'_a})
Y_{lm}(\theta_2,\phi_2)=0
\hspace{10pt} .
\end{eqnarray}
The solution $a^s_{m_a}$ is independent of $m_b$, since the generators
of the binary polyhedral group act only on $|j,m_a\rangle$,
see ansatz (\ref{Eq:Ansatz_red_Basis}).
The set of equations for $a^s_{m_a}$ with $m_a=k N$ and $|m_a|\leq j$,
$k\in \mathbb{Z}$
can be numerically solved using a singular value decomposition routine
by choosing the value $m_b=0$
(see also (E.14) in \cite{Gundermann_2005}).

The eigenmodes of the binary polyhedral spaces are given in terms of the 
spherical basis $\Psi_{jlm}(\tau,\theta,\phi)$ by
\begin{eqnarray}
\label{Eq:eigenmodes_polyhedral_final}
\psi^{\, \cal M}_{j;i}(\tau,\theta,\phi)
\nonumber =
\sum_{l=0}^{2j}\sum_{m=-l}^{l} \xi^{j,i}_{lm}({\cal M})\,
\Psi_{jlm}(\tau,\theta,\phi)
\;\;\\
\xi^{j,i}_{lm}({\cal M})
=\langle jm_ajm_b(i)|lm\rangle a^{s(i)}_{m_a} \\
\nonumber
\hbox{with} \hspace{5pt} m_a + m_b = m  \hspace{5pt} , \hspace{5pt}
m_a\equiv 0 \;\hbox{mod}\;N \hspace{5pt}\hbox{and} 
\hspace{5pt}|m_a|,|m_b|\le j
\hspace{10pt} ,
\end{eqnarray}
where $i$ counts the degenerated modes.
Since these manifolds are homogeneous,
the eigenmodes can be chosen independent of the observer position.
Note, that an additional phase factor $(-\hbox{i})^l$  has to be taken into
account with respect to the radial function of \cite{Abbott_Schaefer_1986}.

\section{Eigenmodes on Polyhedral Double-Action Manifolds}
\label{sec:eigenmodes_polyhedral_da}

In this section the eigenmodes of the Laplace-Beltrami operator on 
the  polyhedral double-action manifolds
${\cal M}=TZ(24,n)$, $OZ(48,n)$, and $IZ(120,n)$, $n>1$, are derived
in terms of  the spherical basis (\ref{Eq:eigenmode_S3_spherical_abstract}).
The action of the cyclic group $Z_n$ onto the 
ansatz\,(\ref{Eq:Ansatz_red_Basis}) does not depend on the action 
of the polyhedral groups.
Therefore, using the ansatz\,(\ref{Eq:Ansatz_red_Basis}),
the eigenmode for the observer position defined by the transformation $t$
is obtained
\begin{eqnarray}
\label{Eq:eigenmodes_TZ_IZ_OZ_1}
\nonumber
\hat{\psi}^{\, \cal M}_{j,i}(\tau,\theta,\phi) & := &
D(t^{-1})\psi^{\, \cal M}_{j,i}(\tau,\theta,\phi) \; = \;
\langle \tau,\theta,\phi|D(t^{-1})|j,i \rangle \\
\nonumber
& = &
\sum_{m_a\equiv 0 \;\hbox{\scriptsize mod}\;N}a^{s(i)}_{m_a}\;
\langle \tau,\theta,\phi|D(t^{-1})|j;m_a,m_b(i)\rangle
\\ \nonumber
& & \hspace{80pt} \hbox{with}
\hspace{5pt}2\,m_b\equiv 0 \;\hbox{mod}\;n
\hspace{10pt} 
\end{eqnarray}
in terms of the spherical coordinates $(\tau,\theta,\phi)$.
Here, the solutions $a^s_{m_a}$ are determined by the set of 
equations (\ref{Eq:equation_polyhedral_final}).
Because of the condition $2\,m_b\equiv 0 \;\hbox{mod}\;n$,
the manifolds with $n>1$ are inhomogeneous. 
For this reason an observer dependence has to be taken into account by the 
translation $D(t^{-1})=e^{\hbox{\scriptsize i}\,(-\alpha + \epsilon)\,J_{bz}}
e^{\hbox{\scriptsize i}\,(-2\rho)\,J_{by}}
e^{\hbox{\scriptsize i}\,(-\alpha -\epsilon)\,J_{bz}}$
which acts only on the states $|j, m_b \rangle$.
Inserting the completeness relation 
$\sum_{lm}|j; l, m \rangle \langle j; l, m|={\bf 1}$, the eigenmodes can be 
rewritten  in terms of the spherical basis $\Psi_{jlm}(\tau,\theta,\phi)$
\begin{equation}
\label{Eq:eigenmodes_TZ_IZ_OZ_2}
\hspace{-30pt}
\hat{\psi}^{\, \cal M}_{j,i}(\tau,\theta,\phi)
=\sum_{lm}\sum_{m_a\equiv 0 \;\hbox{\scriptsize mod}\;N}a^{s(i)}_{m_a}\;
\langle j; l, m|D(t^{-1})|j;m_a,m_b(i)\rangle
\Psi_{jlm}(\tau,\theta,\phi)
\hspace{1pt} . 
\end{equation}
In the next step the completeness relation 
$\sum_{\tilde{m}_b}|j, \tilde{m}_b \rangle \langle j, \tilde{m}_b|={\bf 1}$ and
eq.\,(\ref{Eq:D_function_rho_alpha_epsilon}) are used
to obtain the final expansion
\begin{eqnarray}
\hat{\psi}^{\, \cal M}_{j;i}(\tau,\theta,\phi)
\nonumber & = &
\sum_{l=0}^{2j}\sum_{m=-l}^{l} \xi^{j,i}_{lm}({\cal M};t)\,\Psi_{jlm}(\tau,\theta,\phi)
\\
\label{Eq:eigenmodes_TZ_IZ_OZ_spherical_final}
\xi^{j,i}_{lm}({\cal M};t)
& = &\sum_{\tilde{m}_b}\langle jm_aj\tilde{m}_b|lm\rangle\, a^{s(i)}_{m_a} \,
D^{\,j}_{\tilde{m}_b,m_b(i)}(t^{-1})
\end{eqnarray}
with $m_a + \tilde{m}_b = m$, $m_a\equiv 0\;\hbox{mod}\;N$,
$2\,m_b\equiv 0 \;\hbox{mod}\;n$, and $|m_a|,|m_b|\le j$.
The condition $m_a\equiv 0\;\hbox{mod}\;N$ can be interpreted in such a way
that the coefficients $a^{s(i)}_{m_a}$ vanish for $m_a\neq kN$
with $k\in\mathbb{Z}$.
This expansion corresponds to eq.\,(\ref{Eq:eigenfunction_TZ_IZ_OZ_exp_sph}).
The eigenmodes of the inhomogeneous manifolds ${\cal M}=TZ(24,n)$, $OZ(48,n)$,
$IZ(120,n)$ are expressed in this way by the coefficients $a^{s}_{m_a}$
of the homogeneous spaces $TZ(24,1)$, $OZ(48,1)$, and $IZ(120,1)$,
whose computation is described in \ref{sec:eigenmodes_polyhedral}.

In the numerical evaluation of eq.\,(\ref{Eq:quadratic_sum_da_TZ_OZ_IZ}),
we make use of the invariance $(\alpha-\epsilon)\to-(\alpha-\epsilon)$,
which we would like to derive now.
To that aim consider the transformation of the complex conjugated eigenmode
$\tilde{\psi}^{\, \cal M}_{j,i}(\tau,\theta,\phi) =
D(t^{-1})\big(\psi^{\, \cal M}_{j,i}(\tau,\theta,\phi)\big)^\star$.
The derivation which leads to (\ref{Eq:eigenmodes_TZ_IZ_OZ_spherical_final})
can be repeated for $\tilde{\psi}^{\, \cal M}_{j,i}$ which results in
\begin{eqnarray}
\tilde{\psi}^{\, \cal M}_{j;i}(\tau,\theta,\phi)
\nonumber & = &
\sum_{l=0}^{2j}\sum_{m=-l}^{l} \tilde{\xi}^{j,i}_{lm}({\cal M};t)\,
\Psi_{jlm}(\tau,\theta,\phi)
\\
\label{Eq:eigenmodes_TZ_IZ_OZ_spherical_cc}
\tilde{\xi}^{j,i}_{lm}({\cal M};t)
& = &
(-1)^{l+m} \sum_{\tilde{m}_b}\langle jm_aj\tilde{m}_b|l-m\rangle\,
(a^{s(i)}_{m_a})^\star \, D^{\,j}_{\tilde{m}_b,m_b(i)}(t^{-1})
\hspace{10pt} ,
\end{eqnarray}
where the relation $\Psi_{jlm}^\star(\tau,\theta,\phi) = 
(-1)^{l+m}\Psi_{jl-m}(\tau,\theta,\phi)$ is used.
Here and in the following the same conditions are imposed as stated
below (\ref{Eq:eigenmodes_TZ_IZ_OZ_spherical_final}).
The required sum (\ref{Eq:quadratic_sum_da_TZ_OZ_IZ}) reads now
\begin{eqnarray}
\nonumber
\frac{1}{2l+1}\sum_{m=-l}^{l}& &\sum_{i=1}^{r^{\cal M}(\beta)}
\left|\tilde{\xi}^{j,i}_{lm}({\cal M};t)\right|^2 \\
\nonumber
=
\frac{1}{2l+1}\sum_{m=-l}^{l}& &\sum_{i=1}^{r^{\cal M}(\beta)}
\left|\sum_{\tilde{m}_b}\langle jm_aj\tilde{m}_b|l-m\rangle \,
(a^{s(i)}_{m_a})^\star \,
e^{-\hbox{\scriptsize i}\,\tilde{m}_b\,(\alpha-\epsilon)}\,
d^{\,j}_{\tilde{m}_b, m_b(i)}(-2 \rho)\right|^2
\\ \nonumber
=
\frac{1}{2l+1}\sum_{m=-l}^{l}& &\sum_{i=1}^{r^{\cal M}(\beta)}
\left|\sum_{\tilde{m}_b}\langle jm_aj\tilde{m}_b|lm\rangle \,
a^{s(i)}_{m_a} \,
e^{\hbox{\scriptsize i}\,\tilde{m}_b\,(\alpha-\epsilon)}\,
d^{\,j}_{\tilde{m}_b, m_b(i)}(-2 \rho)\right|^2
\hspace{10pt} ,
\end{eqnarray}
where in the last step the summation index is changed from $-m$ to $m$
and the terms within the modulus are replaced by their complex conjugated
counterparts.
This sum is obtained from the equivalent basis
(\ref{Eq:eigenmodes_TZ_IZ_OZ_spherical_cc}),
and thus refers to the same observer position as the sum
given in (\ref{Eq:quadratic_sum_da_TZ_OZ_IZ}),
which in turn have to be identical.
This can only be achieved if the symmetry
$(\alpha-\epsilon)\to-(\alpha-\epsilon)$ applies
as a comparison with (\ref{Eq:quadratic_sum_da_TZ_OZ_IZ}) shows.


\section*{Acknowledgements}

We would like to thank the Deutsche Forschungsgemeinschaft
for financial support (AU 169/1-1).
The WMAP data from the LAMBDA website (lambda.gsfc.nasa.gov)
were used in this work.


\section*{References}

\bibliography{../bib_astro}

\bibliographystyle{h-physrev5}

\end{document}